\documentclass[aps,pre,twocolumn,showpacs,superscriptaddress,floatfix]{revtex4}
\usepackage{graphicx}
\usepackage{amssymb}
\newcommand{\vd}[2]{\frac{\delta #1}{\delta #2}}
\renewcommand{\k}{{\mathbf k}}
\newcommand{\w}{\omega}
\newcommand{\W}{\Omega}
\newcommand{\x}{{\mathbf x}}
\newcommand{\nkv}[1]{n_{\mathbf #1}}
\newcommand{\Wd}{\Omega^{(d)}}
\newcommand{\dTd}[1]{d \Theta^{(d)}_{\mathbf #1}}
\newcommand{\n}[1]{n_{#1}}
\newcommand{\N}[1]{N_{#1}}
\newcommand{\pd}[2]{\frac{\partial #1}{\partial #2}}
\newcommand{\dd}[2]{\frac{d #1}{d #2}}
\newcommand{\ckz}{c_{\rm KZ}}

\begin{document}

\title{Numerical Solutions of the Isotropic 3-Wave Kinetic Equation}
\author{Colm Connaughton}
\email{connaughtonc@gmail.com} 
\affiliation {Mathematics Institute and Centre for Complexity Science, University of Warwick, Coventry CV4 7AL, UK}

\date{\today}

\begin{abstract}
We show that the isotropic 3-wave kinetic equation is equivalent to the mean field rate 
equations for an aggregation-fragmentation problem with an unusual fragmentation
mechanism. This analogy is used to write the theory of 3-wave 
turbulence almost entirely in terms of a single scaling parameter. A 
new numerical method for solving the kinetic equation over a large
range of frequencies is developed by extending Lee's method for solving
aggregation equations. The new algorithm is validated against some analytic
calculations of the Kolmogorov-Zakharov constant for some families of model
interaction coefficients. The algorithm is then applied to study some 
wave turbulence problems in which  the finiteness of the dissipation scale is
an essential feature. Firstly, it is shown that for finite capacity cascades, 
the dissipation of energy becomes independent of the cut-off frequency as this
cut-off is taken to infinity. This is an explicit indication of the presence 
of a dissipative anomaly. Secondly, a preliminary numerical study is
presented of the so-called bottleneck effect in a wave turbulence context. It is
found that the structure of the bottleneck depends non-trivially on the
interaction coefficient. Finally some results are presented
on the complementary phenomenon of thermalisation in closed wave systems
which demonstrates explicitly for the first time the existence of so-called
mixed solutions of the kinetic equation which exhibit aspects of both
Kolmogorov-Zakharov and equilibrium equipartition spectra.
\end{abstract}

\maketitle

\section{Introduction}
\label{sec-intro}
The 3-wave kinetic equation is the analogue of the Boltzmann equation for
an ensemble of nonlinear dispersive waves interacting weakly via a quadratic
nonlinearity in the wave equation. Such wave systems are often conveniently
modeled using Hamiltonian equations for the complex wave amplitudes, $a_{\k}$
supplemented with additional terms modeling forcing, $f_{\k}$, and 
dissipation, $ \Gamma_{\k}$:
\begin{equation}
\label{eq-HamiltonEqns}
\pd{a_{\k}}{t} = i \vd{H}{\bar{a}_{\k}} + f_{\k} - \Gamma_{\k} a_{\k}.
\end{equation}
The Hamiltonian, $H$, contains quadratic and cubic terms in the wave amplitudes, $a_{\k}$:
\begin{equation}
H = \int \omega_\k a_{\k}\bar{a}_{\k} d\k + \int u(\k) d\k,
\end{equation}
where
\begin{equation}
u(\k_1)\!\!\! =\!\!\! \int L_{\k_1\k_2\k_3}\left( a_{\k_1}a_{\k_2}\bar{a}_{\k_3}+\bar{a}_{\k_1}\bar{a}_{\k_2}a_{\k_3}\right) \delta(\k_1-\k_2-\k_3)\,d\k_2d\k_3.
\end{equation}
The theory of weak wave turbulence \cite{ZLF92,NNB01} studies the statistical evolution of the
solutions of Eq.~(\ref{eq-HamiltonEqns}) in the situation where the nonlinear
term can be treated as a perturbation. Within the framework of weak wave
turbulence, the 3-wave kinetic equation is derived as a consistent asymptotic
closure of the cumulant hierarchy generated by Eq.~(\ref{eq-HamiltonEqns}).
It describes the time evolution of the
spectral wave-action density, $n_{\k}$,  which, for statistically homogeneous
wave fields, is obtained from the two-point correlation function of the
wave amplitudes:
\begin{equation}
\langle  a_{\k_1} \bar{a}_{\k_2}\rangle = n_{\k_1}\,\delta(\k_1-\k_2).
\end{equation}
It takes the form
\begin{equation}
\label{eq-3WKE}
\pd{\nkv{\k_1}}{t} = S[\nkv{k}] +  F[\nkv{k}]  -  D[\nkv{k}].
\end{equation}
$F[\nkv{k}]$, which is absent for decay problems, represents the wave forcing.
$D[\nkv{k}]$ represents dissipation and is typically only present at high
wave-vectors. $S[\nkv{k}]$, referred to as the collision integral, describes
the conservative transfer of energy between wave modes due to resonant
interactions and is the term responsible for the energy cascade. We shall
study its explicit form in Sec.~\ref{sec-3WKE}.

It is convenient, and physically relevant, to consider isotropic scale 
invariant systems.
This means that both the dispersion relation and the nonlinear interaction
coefficient are homogeneous functions of their arguments without any
preferred direction. We denote their degrees of homogeneity by $\alpha$ and 
$\beta$ respectively:
\begin{eqnarray}
\label{eq-homogeneity} \w_{a\,k} &=& a^\alpha\,\w_{\k}\\
\nonumber  L_{a\k_1a\k_2a\k_3} &=& a^\beta\, L_{\k_1\k_2\k_3}.
\end{eqnarray}
For convenience, we shall take the dispersion relation to be a simple power law:
\begin{equation}
\label{eq-dispersionReln} \w{k} = \w{k} = c\,\w^\alpha.
\end{equation}

A huge amount is known about the stationary solutions of Eq.~(\ref{eq-3WKE})
in the turbulent regime where the forcing and dissipation scales are
asymptotically separated from each other in scale. In addition to the 
thermodynamic equilibrium solution, there exists an exact stationary
non-equilibrium solution, known as the Kolmogorov-Zakharov spectrum, which
carries a constant flux of energy through scales. This energy cascade solution
is the analogue of the (phenomenological) Kolmogorov $k^{-5/3}$ spectrum 
of hydrodynamic turbulence. The fact that the energy cascade spectrum can
be derived analytically is one of the principle reasons for theoretical
interest in weak wave turbulence.

Rather less is known about the solutions of Eq.~(\ref{eq-3WKE}) beyond the 
characterisation of the stationary state in the limit where the forcing
and dissipation scales tend to zero and infinity respectively. In particular,
knowledge about the dynamical evolution of the solutions is restricted to
a subset of systems for which a self-similar solution can be constructed
using energy conservation arguments which in any case, leave the scaling
function undetermined. Similarly relatively little is known about how the 
system matches itself to the source and sink in the case of  finite forcing 
and dissipation scales which break the scale invariance necessary to obtain
the K-Z solution. It is in this context that the present work fits.

This article contains two main ideas. The first is that in the case of
isotropic systems, the 3-wave kinetic equation is equivalent to the rate
equations for an aggregation-fragmentation problem with a rather unusual 
fragmentation process. This is useful for several reasons. Firstly,
 there is a large body of knowledge about rate equations for 
aggregation-fragmentation problems which might provide useful insights.
Secondly, this description is very compact with almost all properties of
the solution being determined by a single scaling parameter. Thirdly,
and this forms the basis for the second main idea of the article, this
description forms the basis for a new numerical procedure for solving
the 3-wave kinetic equation which can resolve a very large range of
scales compared to a direct numerical integration. This numerical procedure
can then be used to investigate aspects of 3-wave turbulence which are less
amenable to analytic understanding.

The layout of the article is as follows. We first explain in Sec.~\ref{sec-3WKE}the analogy between 3-wave turbulence and aggregation-fragmentation equations,
with the technical details relegated to an appendix. One of the principle
insights provided by this analogy is that most of the scaling properties of
the system are determined by a single parameter, 
$\lambda=\frac{2\beta-\alpha}{\alpha}$. In Sec.~\ref{sec-WTResults} we devote some
time to explaining how the standard results of weak wave turbulence are
expressed in terms of $\lambda$. In Sec.~\ref{sec-truncation} we consider
truncating the system at some finite frequency and discuss open and
closed truncations, two distinct natural choices of truncation. In 
Sec.~\ref{sec-numerics} the results of Sec.~\ref{sec-3WKE} and 
Sec.~\ref{sec-truncation} are used to develop a new numerical procedure which
allows the stable integration of the 3-wave kinetic equation over many
decades of frequencies. Some technical details of the method are postponed
to a second appendix. The remainder of the article is then devoted to
presenting some preliminary studies which are intended to demonstrate the
usefulness of this algorithm. In Sec.~\ref{sec-dissipativeAnomaly} we
study the numerical signature of the dissipative anomaly in finite capacity
cascades. In Sec.~\ref{sec-bottleneck} we demonstrate the non-trivial 
structure of the bottleneck effect in the 3-wave kinetic equation with
an open truncation and in Sec.~\ref{sec-thermalisation} we study the
thermalisation phenomenon which occurs when the equation is subjected to
a closed truncation. The article closes with some conclusions and 
speculations about future directions of research.

\section{Formulation of the Isotropic 3-Wave Kinetic Equation as an Aggregation--Fragmentation Problem}
\label{sec-3WKE}
The collision 
integral is usually written in the form 
\begin{equation}
\label{eq-S}
 S[\nkv{k}] = \int_{{\mathbf R}^{2d}} \left( R_{\k_1\k_2\k_3} - R_{\k_2\k_3\k_1} -  R_{\k_3\k_1\k_2}\right)\,d\k_2d\k_2
\end{equation}
where
\begin{eqnarray}
\nonumber R_{\k_1\k_2\k_3} &=& 4\pi\,  L_{\k_1\k_2\k_3}^2 (n_{\k_2} n_{\k_3} -  n_{\k_1}n_{\k_3}- n_{\k_1}n_{\k_2}) \\
\nonumber && \hspace{0.5cm} \delta(\omega_{\k_1}-\omega_{\k_2}-\omega_{\k_3})\, \delta(\k_1-\k_2-\k_3)
\end{eqnarray}
The total wave-action, $N$, and total quadratic energy, $E$, in the system are
\begin{eqnarray}
\label{eq-NDefn1} N  &=& \int_{{\mathbf R}^{d}} \nkv{k_1}\,d\k_1\\
\label{eq-EDefn1} E  &=& \int_{{\mathbf R}^{d}} \w_{\k_1}\,\nkv{k_1}\,d\k_1
\end{eqnarray}
respectively. $E$ is conserved by Eq.~(\ref{eq-3WKE}) in the absence of forcing
and dissipation. For isotropic systems it is convenient to work with
the angle-averaged frequency spectrum, $\N{\w}$ instead of the basic $\k$-space
spectrum, $\nkv{k}$. $\N{\w}$ is defined such that 
$\int_{\w_1}^{\w_2} \N{\w}\,d\w$ is the total wave action in the frequency band
$[\w_1,\w_2]$. It is shown in the appendix that, for isotropic systems,
Eq.~(\ref{eq-3WKE}) is equivalent to
\begin{equation}
\label{eq-3WKEB}
\pd{\N{\w_1}}{t} = S_1[\N{\w}] + S_2[\N{\w}] + S_3[\N{\w}] + F[\N{\w}] -  D[\N{\w}]
\end{equation}
where
\begin{eqnarray}
\nonumber S_1[\N{\w}]\!\!\! &=& \!\!\!\!\!\int\!\! K_1(\w_2,\w_3)\, \N{\w_2} \N{\w_3} \delta(\w_1\!-\!\w_2\!-\!\w_3)\, d\w_{23}\\
\label{eq-S1} &-& \!\!\!\!\!\int\!\! K_1(\w_3,\w_1)\, \N{\w_1} \N{\w_3} \delta(\w_2\!-\!\w_3\!-\!\w_1)\, d\w_{23}\\
\nonumber  &-& \!\!\!\!\!\int\!\! K_1(\w_1,\w_2)\,\N{\w_1} \N{\w_2} \delta(\w_3\!-\!\w_1\!-\!\w_2)\, d\w_{23},
\end{eqnarray}
\begin{eqnarray}
\nonumber S_2[\N{\w}]=\!\!\! &-& \!\!\!\!\!\int\!\! K_2(\w_2,\w_3)\, \N{\w_1} \N{\w_2} \delta(\w_1\!-\!\w_2\!-\!\w_3)\, d\w_{23}\\
\label{eq-S2} &+& \!\!\!\!\!\int\!\! K_2(\w_3,\w_1)\, \N{\w_2} \N{\w_3} \delta(\w_2\!-\!\w_3\!-\!\w_1)\, d\w_{23}\\
\nonumber  &+& \!\!\!\!\!\int\!\! K_2(\w_1,\w_2)\,\N{\w_1} \N{\w_3} \delta(\w_3\!-\!\w_1\!-\!\w_2)\, d\w_{23}
\end{eqnarray}
and
\begin{eqnarray}
\nonumber S_3[\N{\w}]=\!\!\! &-& \!\!\!\!\!\int\!\! K_3(\w_2,\w_3)\, \N{\w_1} \N{\w_3} \delta(\w_1\!-\!\w_2\!-\!\w_3)\, d\w_{23}\\
\label{eq-S3} &+& \!\!\!\!\!\int\!\! K_3(\w_3,\w_1)\, \N{\w_1} \N{\w_2} \delta(\w_2\!-\!\w_3\!-\!\w_1)\, d\w_{23}\\
\nonumber  &+& \!\!\!\!\!\int\!\! K_3(\w_1,\w_2)\,\N{\w_2} \N{\w_3} \delta(\w_3\!-\!\w_1\!-\!\w_2)\, d\w_{23}.
\end{eqnarray}
$F[\N{\w}]$ and $D[\N{\w}]$ are forcing and dissipation terms whose exact forms
depend on the problem under consideration.
In these formulae, as shown in the appendix, $K_1(\w_i,\w_j)$,  
$K_2(\w_i,\w_j)$ and  $K_3(\w_i,\w_j)$ are homogeneous functions which an be
constructed from the original interaction coefficient, $L_{\k_1\k_2\k_3}$.
They all have degree of homogeneity
\begin{equation}
\label{eq-lambdaDefn}
\lambda=\frac{2\beta - \alpha}{\alpha}.
\end{equation}
We have therefore already obtained a nontrivial result which seems to
have gone unnoticed before: all scaling properties of Eq.~(\ref{eq-3WKE}),
which contains 3 scaling parameters, $\alpha$,$\beta$
and $d$, seem to depend on a single scaling parameter, $\lambda$ given by
Eq.~(\ref{eq-lambdaDefn}). In fact, this is not completely true. Owing to
the fact that $K_1(\w_i,\w_j)$, $K_2(\w_i,\w_j)$ and  $K_3(\w_i,\w_j)$ are
not identical functions (even though they have the same degree of
homogeneity), some memory of $\alpha$,$\beta$ and $d$ is retained in
their internal structure - see Eq.~(\ref{eq-defnK2K3}) - which can sometimes
be important. The details of what can and cannot be expressed solely in terms
of the new parameter, $\lambda$, will be addressed in Sec.~\ref{sec-WTResults}.

\begin{figure}
\begin{tabular}{l}
({\bf A}) : $S_1[\N{\w}]$:\\
\includegraphics[width=6.5cm]{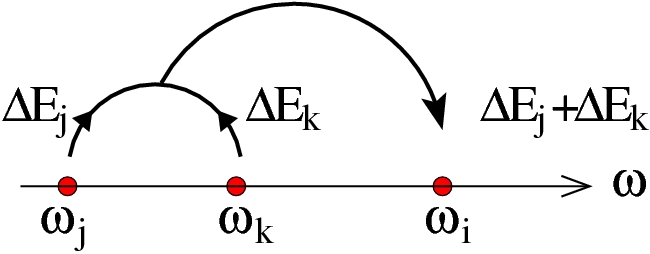}\\
({\bf B}) : $S_2[\N{\w}]$:\\
\includegraphics[width=6.5cm]{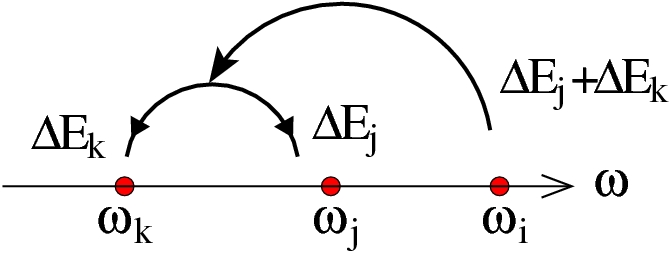}\\
({\bf C}) : $S_3[\N{\w}]$:\\
\includegraphics[width=6.5cm]{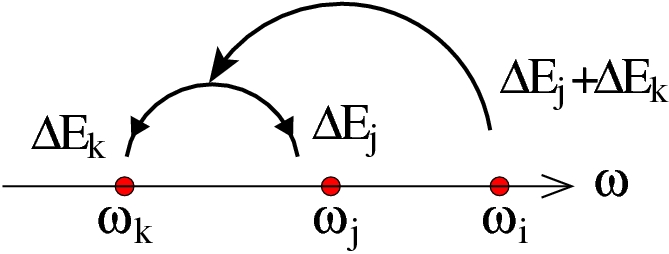}
\end{tabular}
\caption{\label{fig-S} Graphical representations of the dynamics encoded
by Eq.~(\ref{eq-S1}), Eq.~(\ref{eq-S2}) and Eq.~(\ref{eq-S3}).}
\end{figure}

Let us first investigate the physical meaning of Eq.~(\ref{eq-S1}) --
Eq.~(\ref{eq-S3}). Firstly, it is easily shown that each of the three
collision integrals, $S_1[\N{\w}]$, $S_2[\N{\w}]$ and $S_3[\N{\w}]$,
individually conserve the total energy. 

Let us first consider  Eq.~(\ref{eq-S1}) in isolation. The resulting
kinetic equation
\begin{equation}
\label{eq-SKE}
\pd{\N{\w_1}}{t} = S_1[\N{\w}],
\end{equation}
is actually the Smoluchowski kinetic equation \cite{LEY2003} which describes
the mean-field dynamics of cluster--cluster aggregation although it has
been written in a somewhat non-standard form \cite{CRZ2004,CRZ2009}. We can
therefore make an analogy between wave turbulence and cluster--cluster 
aggregation.  In this analogy, wave frequency, $\w$, is analogous to cluster mass, $m$, spectral wave-action density, $\N{\w}$, is analogous to density of
clusters having mass $m$, $\N{m}$ and the wave interaction coefficient,
$K_1(\w_1,\w_2)$ is analogous to the coagulation kernel, $K(m_1,m_2)$.

The physical meaning of the collision integral is easy to understand in the
aggregation context.  In coagulation, pairs of clusters having masses $m_j$ and
$m_k$ aggregate to produce a single cluster having
mass $m_i=m_j+m_k$. The rate at which this process occurs, as for any chemical
rate equation, is proportional to the
aggregation kernel, $K(m_j,m_k)$, and to the densities of clusters having
masses, $m_j$ and $m_k$. This can be thought of an exchange of mass within the 
mass triad, $(m_i, m_j, m_k)$. The mass contained in clusters of mass $m_j$
{\em decreases} at a rate $m_j K(m_j,m_k) \N{m_j} \N{m_k}$, the mass contained
in clusters of mass $m_k$ {\em decreases} at a rate $m_k K(m_j,m_k) \N{m_j} \N{m_k}$ while the mass contained in clusters of mass $m_i$ {\em increases}
at a rate $(m_j+m_k)\,K(m_j,m_k)\, \N{m_j}\, \N{m_k}$. Clearly mass is
conserved overall. The collision integral therefore simply calculates the net 
rate of change of the density of clusters of a given mass by summing the 
contributions of this process over all mass triads.  The signs of the terms
composing the collision integral now make perfect sense: the first (positive)
term accounts for the rate of increase of clusters of mass $m_1$ due to
the aggregation of pairs of smaller clusters having masses $m_2$ and $m_3$ which
satisfy $m_2+m_3=m_1$. The second and third (negative) terms account for the
rate of decrease of clusters of mass $m_1$ which occurs when such clusters
meet any other cluster having mass $m_2$ or $m_3$ and aggregate to produce
a heavier cluster having mass $m_1+m_2$ of $m_1+m_3$. Of course, these
two negative terms are usually combined into a single term since they differ
only by the labeling of dummy variables. Nevertheless, for reasons which
will become apparent in Sec.~\ref{sec-WTResults} we choose to keep the rather
verbose form of Eq.~(\ref{eq-S1}).

Taking this over to the wave analogy, we see that the first term on the RHS of
Eq.~(\ref{eq-3WKE}) describes a transfer of energy in resonant triads, 
$(\w_i, \w_j,\w_k)$. That is to say, triads for which $\w_i = \w_j+\w_k$. 
The energy contained in waves of frequency $\w_j$
{\em decreases} at a rate $\w_j K_1(\w_j,\w_k) \N{\w_j} \N{\w_k}$, the energy 
contained in waves of frequency $\w_k$ {\em decreases} at a rate 
$\w_k K_1(\w_j,\w_k) \N{\w_j} \N{\w_k}$ while the energy contained in waves of 
frequency $\w_i$ {\em increases}
at a rate $(\w_j+\w_k)\,K_1(\w_j,\w_k)\, \N{\w_j}\, \N{\w_k}$. The rates of
energy transfer for each mode in the triad $(\w_i, \w_j,\w_k)$ can thus be 
summarised as
\begin{eqnarray}
\nonumber \w_j&:& \Delta E_j = -\w_j K_1(w_j,w_k) \N{\w_j}  \N{\w_k}\\
\label{eq-S1Rates} \w_k&:& \Delta E_k = -\w_k K_1(w_j,w_k) \N{\w_j}  \N{\w_k}\\
\nonumber \w_i&:& \Delta E_i = + (\Delta E_j +  \Delta E_k).
\end{eqnarray}

This process is
illustrated graphically in Fig.~\ref{fig-S}(A).  The collision integral 
then calculates the net rate of change of the energy of the waves of each frequency by summing the
contributions of this process over all resonant triads. It is clear from
this discussion that energy can only be transferred from lower 
frequencies to higher frequencies by the collision integral $S_1[\N{\w}]$.
As we shall see below, the reverse is the case for the integrals $S_2[\N{\w}]$
and $S_3[\N{\w}]$. $S_1[\N{\w}]$ is therefore the driver of the direct
cascade in wave turbulence.

If we have laboured the point a little on the interpretation of $S_1[\N{\w}]$
and the analogy with aggregation, it is because the corresponding interpretation
of $S_2[\N{\w}]$ and $S_3[\N{\w}]$ is less obvious. Indeed, the dynamics encoded
by these integrals illustrates clearly why wave resonances are different
from particles. We first notice that the sign structure of $S_2[\N{\w}]$ and 
$S_3[\N{\w}]$ is different from that of $S_1[\N{\w}]$. Two terms are positive
and one is negative. Looking at the negative term, it is clear that the frequency, 
$\w_1$, which is losing energy is the sum of two lower frequencies, $\w_2$
and $\w_3$. Thus the correct pictures to draw for these processes are those
in Fig.~\ref{fig-S}(B) and Fig.~\ref{fig-S}(C): a high frequency mode loses
energy to a pair of lower frequency ones, the reverse of the process
encoded by $S_1[\N{\w}]$. The difference between the two processes is
in the rates and is summarised as follows. For each resonant triad,
$(\w_i, \w_j,\w_k)$ satisfying $\w_i=\w_j+\w_k$, in $S_2[\N{\w}]$ the
rates of energy transfer are:
\begin{eqnarray}
\nonumber \w_j&:& \Delta E_j = +\w_j K_2(w_j,w_k) \N{\w_j}  \N{\w_i}\\
\label{eq-S2Rates}\w_k&:& \Delta E_k = +\w_k K_2(w_j,w_k) \N{\w_j}  \N{\w_i}\\
\nonumber\w_i&:& \Delta E_i = - (\Delta E_j +  \Delta E_k)
\end{eqnarray}
On the other hand, for each resonant triad,
$(\w_i, \w_j,\w_k)$ satisfying $\w_i=\w_j+\w_k$, in $S_3[\N{\w}]$ the
rates of energy transfer are:
\begin{eqnarray}
\nonumber \w_j&:& \Delta E_j = +\w_j K_3(w_j,w_k) \N{\w_k}  \N{\w_i}\\
\label{eq-S3Rates}\w_k&:& \Delta E_k = +\w_k K_3(w_j,w_k) \N{\w_k}  \N{\w_i}\\
\nonumber \w_i&:& \Delta E_i = - (\Delta E_j +  \Delta E_k)
\end{eqnarray}
It is clear from Eqs.~(\ref{eq-S2Rates}) and Eqs.~(\ref{eq-S3Rates}) that
energy can  only be transferred from higher frequencies to lower frequencies
by the collision integrals $S_2[\N{\w}]$ and $S_3[\N{\w}]$. Thus they 
describe back-scatter in the wave turbulent cascade. In the aggregation
analogy, they can be thought of as describing some kind of nonlinear
fragmentation process in which the rate of fragmentation of clusters of 
a given size is proportional to the density of clusters of that size
 and to the density of fragments. This latter dependence makes this a 
rather unusual process from the point of view of interacting particle
systems. Fragmentation is often modeled as a linear process. For a review of
fragmentation see \cite{RED1990}. Although some non-linear models have been 
studied (see, for example, \cite{KB2003} and the references therein) the
fragmentation mechanism at work here is, to the best of our knowledge, new.
The idea that nonlinear fragmentation has some connection to energy 
transfer in turbulence is not a new idea \cite{STEP1989,BK2008} but this
is the first case for which the fragmentation equations can be derived
from the underlying dynamical equation.

In what follows, we shall typically work with simplified model interaction
kernel rather than the complicated functions which would arise from
particular examples of physical wave systems. We introduce the following
model kernel, which has been very extensively studied 
\cite{ERN1986,EVD1988} in the context of cluster--cluster aggregation:
\begin{equation}
\label{eq-modelKernel}
K_1(\w_1,\w_2) = \frac{1}{2}\left(\w_1^\mu\w_2^\nu + \w_1^\nu\w_2^\mu \right).
\end{equation}
Here the exponents $\mu$ and $\nu$ must satisfy $\mu+\nu=\lambda$. Two 
special cases are of particular interest. The first is the product kernel:
\begin{equation}
\label{eq-productKernel}
K_1(\w_1,\w_2) = \left(\w_1\w_2\right)^{\frac{\lambda}{2}}.
\end{equation}
The second is the sum kernel:
\begin{equation}
\label{eq-sumKernel}
K_1(\w_1,\w_2) =  \frac{1}{2}\left(\w_1^\lambda + \w_2^\lambda \right).
\end{equation}

In developing the analogy between 3-wave turbulence and 
aggregation--fragmentation problems, it is worth pointing out that, in the
context if aggregation, it is common to work with a discrete analogue of 
Eq.~(\ref{eq-SKE}). This is for relevant for the case of so-called mono-disperse
initial conditions which means that all particles initially have the same
mass which can be taken equal to one. The dynamics will only produce 
clusters having integer masses and Eq.~(\ref{eq-SKE}) can then be presented
as an infinite set of coupled ordinary differential equations for the mass
densities, $N_i$, with the integrals having been replaced by sums. A
discrete point of view is relevant for the wave kinetic equation too. In decay 
problems with monochromatic initial conditions or forced problems with 
monochromatic forcing, it is clear from Eq.~(\ref{eq-3WKEB}) that only 
multiples of the input frequency, $\w_0$, can be excited by the dynamics. 
Taking $\w_0=1$, the discrete frequencies $\w_i$, are then integers, and
replacing the integrals with sums, we obtain the following infinite set 
of coupled ordinary differential equations describing the time evolution
of the infinite vector of discrete wave 
occupation numbers, ${\bf N}(t)=(N_1(t), N_2(t),\ldots)$:
\begin{equation}
\label{eq-discreteKE}\dd{N_i}{t} = S_i[{\bf N}]  +  F_i[{\bf N}] -  D_i[{\bf N}]\ \ \ \ \mbox{$i=1,2,\ldots$}
\end{equation}
where
\begin{displaymath}
S_i[{\bf N}]  = \,S_i^{(1)}[{\bf N}] +  \,S_i^{(2)}[{\bf N}] +  \,S_i^{(3)}[{\bf N}]
\end{displaymath}
with the discrete collision integrals given by
\begin{eqnarray}
\nonumber S_i^{(1)}[{\bf N}] =& & \sum_{j=1}^{i-1} K_1(j,i-j)\, N_j N_{i-j}\\
\label{eq-discreteS1} &-& \sum_{j=i+1}^{\infty} K_1(j-i,i)\, N_i N_{j-i}\\
\nonumber  &-&  \sum_{j=1}^{\infty} K_1(i,j)\,N_i N_{j},
\end{eqnarray}
\begin{eqnarray}
\nonumber S_i^{(2)}[{\bf N}] = &-& \sum_{j=1}^{i-1} K_2(j,i-j)\, N_i N_{j}\\
\label{eq-discreteS2} &+& \sum_{j=i+1}^{\infty} K_2(j-i,i)\, N_j N_{j-i} \\
\nonumber  &+& \sum_{j=1}^{\infty} K_2(i,j)\,N_i N_{i+j}
\end{eqnarray}
and
\begin{eqnarray}
\nonumber S_i^{(3)}[{\bf N}] =&-& \sum_{j=1}^{i-1} K_3(j,i-j)\, N_i N_{i-j}\\
\label{eq-discreteS3} &+& \sum_{j=i+1}^{\infty} K_3(j-i,i)\, N_j N_{i} \\
\nonumber  &+& \sum_{j=1}^{\infty} K_3(i,j)\,N_j N_{i+j}
\end{eqnarray}
The forcing and dissipation terms should be chosen appropriately according to
the problem under study. In this section we have shown that it is possible to
think of 3-wave turbulence from the perspective of aggregation-fragmentation
problems. In the remainder of the article, we shall demonstrate the usefulness
of this analogy for understanding turbulence. The analogy turns out to be 
useful in both directions however. For a discussion of stochastic aggregation
from the point of view of turbulence theory, see \cite{CRZ2005,CRZ2006}.

\section{A Review of Some Standard Wave Turbulence Results}
\label{sec-WTResults}
In this section we write down some of the standard results of wave turbulence
theory in the language of the previous section. 

\subsection{Kolmogorov-Zakharov Spectrum}
One of the principal results of wave turbulence theory is the fact that, for scale
invariant systems, the kinetic equation has an exact stationary solution
which carries a constant flux of energy through scales. This solution, 
known as the Kolmogorov-Zakharov (KZ) spectrum, is the direct analogue for
waves of the well known Kolmogorov $k^{-5/3}$ spectrum characteristic of
hydrodynamic turbulence. For isotropic systems, the KZ spectrum is usually
presented as a stationary solution of Eq.~(\ref{eq-3WKE}) in the limit
where the forcing wave-number tends to zero and the dissipation wave-number 
tends to infinity. It is usually written \cite{ZLF92,NNB01}:
\begin{equation}
\label{eq-KZSpectrumk}
\nkv{\k} = \ckz^{(k)}\,k^{-\beta-d}
\end{equation}
where $\ckz^{(k)}$ is a dimensional constant which can be calculated.

The key step in obtaining the KZ spectrum from Eq.~(\ref{eq-3WKE}) is to apply 
a change of variables known as the Zakharov transformation to the second
and third integrals in Eq.~(\ref{eq-3WKE}). The idea of the transformation
is to map the supports of the frequency delta functions in the second and
third integrals onto that of the first which allows the stationary
solution to be clearly seen. The Zakharov transformation can be applied
individually to each of the collision integrals in Eq.~(\ref{eq-3WKEB}). We
shall demonstrate the procedure explicitly for $S_1[\N{\w}]$ and then 
write down the analogous results for $S_2[\N{\w}]$ and $S_3[\N{\w}]$.

Let us seek a solution of the form $\N{\w} = \ckz\,\w^{-x}$ with $x$ to
be determined so that
\begin{eqnarray}
\nonumber 0\!\!\! &=& \!\!\!\!\!\ckz^2\int\!\! K_1(\w_2,\w_3)\, (\w_2\w_3)^{-x} \delta(\w_1\!-\!\w_2\!-\!\w_3)\, d\w_{23}\\
\label{eq-S1C} &-& \!\!\!\!\!\ckz^2\int\!\! K_1(\w_3,\w_1)\, (\w_1\w_3)^{-x} \delta(\w_2\!-\!\w_3\!-\!\w_1)\, d\w_{23}\\
\nonumber  &-& \!\!\!\!\!\ckz^2\int\!\! K_1(\w_1,\w_2)\,(\w_1\w_2)^{-x} \delta(\w_3\!-\!\w_1\!-\!\w_2)\, d\w_{23}.
\end{eqnarray}
We now apply the following changes of variables:
\begin{equation}
\label{eq-ZT1}(\w_2, \w_3) \to \left(\frac{\w_1^2}{\w_2}, \frac{\w_1\w_3}{\w_2} \right)
\end{equation}
and
\begin{equation}
\label{eq-ZT2}(\w_2, \w_3) \to \left(\frac{\w_1\w_2}{\w_3}, \frac{\w_1^2}{\w_3} \right)
\end{equation}
to the second and third integrals in Eq.~(\ref{eq-S1C}) respectively. Noting 
that the respective Jacobians are $\left(\frac{\w_1}{\w_2}\right)^3$ and 
$\left(\frac{\w_1}{\w_3}\right)^3$ and utilising the fact that $K_1(\w_2,\w_3)$
is a homogeneous function of degree $\lambda$, some algebra yields the 
following:
\begin{eqnarray}
\nonumber 0\!\!\! &=& \!\!\!\!\!\ckz^2\int\!\! K_1(\w_2,\w_3)\, (\w_2\w_3)^{-x} \delta(\w_1\!-\!\w_2\!-\!\w_3)\, d\w_{23}\\
\label{eq-S1D} &-& \!\!\!\!\!\ckz^2\int\!\! K_1(\w_2,\w_3)\, \left(\frac{\w_1}{\w_2}\right)^{\lambda+2-2x}(\w_2\w_3)^{-x}\\
\nonumber & &\hspace{3.5cm} \delta(\w_1\!-\!\w_2\!-\!\w_3)\, d\w_{23}\\
\nonumber  &-& \!\!\!\!\!\ckz^2\int\!\! K_1(\w_2,\w_3)\, \left(\frac{\w_1}{\w_3}\right)^{\lambda+2-2x}(\w_2\w_3)^{-x}\\
\nonumber & &\hspace{3.5cm}  \delta(\w_1\!-\!\w_2\!-\!\w_3)\, d\w_{23}.
\end{eqnarray}
This can be put into a single integral,
\begin{eqnarray}
\label{eq-S1AfterZT} 0\!\!\! &=& \!\!\!\!\!\ckz^2\int\!\! K_1(\w_2,\w_3)\, (\w_2\w_3)^{-x} \delta(\w_1\!-\!\w_2\!-\!\w_3)\\
\nonumber & &\w_1^{\lambda+2-2x}\left[ \w_1^{2x-\lambda-2}-\w_2^{2x-\lambda-2}-\w_3^{2x-\lambda-2}\right]\, d\w_{23},
\end{eqnarray}
from which it is easy to see that the right hand side vanishes when
$2x-\lambda-2 = 1$. This yields the KZ exponent
$x = \frac{\lambda+3}{2}.$
The stationary angle-averaged frequency spectrum is therefore
\begin{equation}
\label{eq-KZSpectrumw}
\N{\w} = \ckz\,\w^{-\frac{\lambda+3}{2}}.
\end{equation}
Using  Eq.~(\ref{eq-transformToFreq}) and Eq.~(\ref{eq-dispersionReln}), it
is clear that Eq.~(\ref{eq-KZSpectrumw}) is equivalent to the more usual
expression for the KZ spectrum given in Eq.~(\ref{eq-KZSpectrumk}). As is
often remarked, Eq.~(\ref{eq-KZSpectrumw}) can also be obtained simply
by dimensional analysis. The true worth of the Zakharov transformations
lies in the fact that they provide a means to obtain the numerical value of
the constant $\ckz$ and to study the conditions under which the spectrum
given by Eq.~(\ref{eq-KZSpectrumw}) is an admissable stationary solution of
the kinetic equation. Finally, returning to the analogy with aggregation,
Eq.~(\ref{eq-KZSpectrumw}) is also well known 
\cite{HEZ1983,HAK1987,KON01,CRZ2004} in the aggregation literature as the  
stationary solution of the Smoluchowski equation in the presence of a source 
of monomers.

Exactly the same steps may be applied to $S_2[\N{\w}]$ and $S_3[\N{\w}]$.
The resulting integrals, analogous to Eq.~(\ref{eq-S1AfterZT}) are
\begin{eqnarray}
\label{eq-S2AfterZT} 0\!\!\! &=& \!\!\!\!\!-\ckz^2\int\!\! K_2(\w_3,\w_2)\, (\w_1\w_3)^{-x} \delta(\w_1\!-\!\w_2\!-\!\w_3)\\
\nonumber & &\w_1^{\lambda+2-2x}\left[ \w_1^{2x-\lambda-2}-\w_2^{2x-\lambda-2}-\w_3^{2x-\lambda-2}\right]\, d\w_{23},
\end{eqnarray}
and
\begin{eqnarray}
\label{eq-S3AfterZT} 0\!\!\! &=& \!\!\!\!\!-\ckz^2\int\!\! K_3(\w_3,\w_2)\, (\w_1\w_2)^{-x} \delta(\w_1\!-\!\w_2\!-\!\w_3)\\
\nonumber & &\w_1^{\lambda+2-2x}\left[ \w_1^{2x-\lambda-2}-\w_2^{2x-\lambda-2}-\w_3^{2x-\lambda-2}\right]\, d\w_{23}
\end{eqnarray}
respectively.

\subsection{Finite and Infinite Capacity Cascades}

The direct energy cascade has {\em infinite capacity} \cite{BNN2001,NNB01} if 
the energy contained in the KZ spectrum diverges at the high-$k$ end and 
{\em finite capacity} otherwise. The notion of capacity is important because
for finite capacity systems forced with a constant energy injection rate, the
cascade necessarily propagates to $k=\infty$ in finite time \cite{FS1991}.
In the usual notation, the KZ spectrum Eq.~(\ref{eq-KZSpectrumk}) has
finite capacity when $\beta > \alpha$. In the notation of Sec.~\ref{sec-3WKE},
the capacity criterion is be determined by considering the energy contained in 
the KZ-spectrum, Eq.~(\ref{eq-KZSpectrumw}) in the range of frequencies
$[\w_0,\W]$:
\begin{eqnarray}
\int_{\w_0}^\W \w \N{\w} d\w &=& \ckz\int_{\w_0}^\W \w^\frac{-\frac{\lambda+1}{2}}d\w\\
&=&  \frac{2\ckz}{1-\lambda}\left[ \W^{\frac{1-\lambda}{2}} - {\w_0}^{\frac{1-\lambda}{2}} \right].
\end{eqnarray}
Looking at what happens as $\W\to\infty$, we see that the cascade has finite 
capacity for $\lambda>1$. In the aggregation analogy, the criterion $\lambda>1$
is known as the condition for the presence of a gelation transition in the
system.

\subsection{Breakdown Criterion and the Generalised Phillips Spectrum}

The derivation of Eq.~(\ref{eq-3WKE}) requires that the linear timescale, 
$\tau_{\rm L}$, associated with the waves is much faster than the nonlinear 
timescale, $\tau_{\rm NL}$, associated with resonant energy transfer between 
waves. This condition may be invalidated by the KZ spectrum, either at
large or small scales \cite{KIT1983,BNN2001,NNB01}, a situation referred to 
as ``breakdown''. The breakdown criterion is derived as follows.
The linear timescale can be estimated
as $\tau_{\rm L}\sim\w^{-1}$. The nonlinear timescale can be estimated as
$\tau_{\rm NL}^{-1} \sim \frac{1}{\N{\w}}\pd{\N{\w}}{t}$. Therefore, on
an arbitrary spectrum, $\N{\w}\sim \w^{-x}$, the ratio 
$\tau_{\rm L}/\tau_{\rm NL}$ can be estimated from Eq.~(\ref{eq-3WKEB}):
\begin{equation}
\label{eq-tLOvertNL} \frac{\tau_{\rm L}}{\tau_{\rm NL}} \sim \w^{\lambda-x}.
\end{equation}
If $x$ is the KZ exponent given by Eq.~(\ref{eq-KZSpectrumw}), then this
ratio becomes
\begin{equation}
 \frac{\tau_{\rm L}}{\tau_{\rm NL}} \sim \w^\frac{\lambda-3}{2}
\end{equation}
from which we conclude that the KZ spectrum breaks down at high frequencies
if $\lambda>3$. Using Eq.~(\ref{eq-lambdaDefn}) to translate this back into
the usual notation, we recover the usual criterion \cite{BNN2001,NNB01} for
breakdown at small scales, $\beta>2\alpha$.

The Generalised Phillips Spectrum \cite{CON2002,NZ2008} is the spectrum for
which the ratio $\tau_{\rm L}/\tau_{\rm NL}$ is independent of the scale. It
is important since it is a likely candidate to replace the KZ spectrum
after breakdown occurs \cite{NZ1992} . From Eq.~(\ref{eq-tLOvertNL}), it is clear that
the Generalised Phillips Spectrum in the notation of Sec.~\ref{sec-3WKE} is
simply
\begin{equation}
\label{eq-GPSpectrum}
\N{\w} \sim \w^{-\lambda}.
\end{equation}
If this spectrum is translated back into the usual notation using
 Eq.~(\ref{eq-transformToFreq}) and Eq.~(\ref{eq-dispersionReln}) we
obtain $\n{k} \sim k^{-(2\beta-2\alpha+d)}$, which is the analogue for
3-wave interactions of the better known formula for the 4-wave case, 
$\n{k} \sim k^{-(\gamma-\alpha+d)}$ \cite{CON2002,NZ2008} (the general formula for $N$-wave
interactions is $\n{k} \sim k^{-(2\gamma_N-2\alpha+(N-2)d)/(N-2)}$).

\subsection{Thermodynamic Spectrum}

In the above application of the Zakharov transformations to the collision
integrals, $S_1[\N{\w}]$, $S_2[\N{\w}]$ and $S_3[\N{\w}]$, we picked out
the KZ spectrum as a stationary solution in each case. We saw no sign of the 
other stationary solution of Eq.~(\ref{eq-3WKE}), the thermodynamic spectrum 
corresponding to equipartition of energy. This is because the equilibrium
spectrum satisfies detailed balance in the sense that the forward and backward 
transfer terms balance each other scale by scale. It is not a property of 
$S_1[\N{\w}]$, $S_2[\N{\w}]$ or $S_3[\N{\w}]$ individually but rather of
the full collision integral. To see this, let us add together 
Eq.~(\ref{eq-S1AfterZT}), Eq.~(\ref{eq-S2AfterZT}) and Eq.~(\ref{eq-S3AfterZT}).
The result is
\begin{widetext}
\begin{eqnarray}
\label{eq-applyZT1}0\!\!\! &=& \!\!\!\ckz^2\int\!\! \left[ K_1(\w_3,\w_2)\, (\w_2\w_3)^{-x} -K_2(\w_3,\w_2)\, (\w_1\w_3)^{-x} -K_3(\w_3,\w_2)\, (\w_1\w_2)^{-x}\right]\\
\nonumber & &\w_1^{\lambda+2-2x}\left[ \w_1^{2x-\lambda-2}-\w_2^{2x-\lambda-2}-\w_3^{2x-\lambda-2}\right]\,  \delta(\w_1\!-\!\w_2\!-\!\w_3)\, d\w_{23}\\
\label{eq-applyZT2}&=& \!\!\!\ckz^2\int\!\! K_1(\w_3,\w_2) \left[(\w_2\w_3)^{-x} -\left(\frac{\w_1}{\w_2}\right)^\frac{\alpha-d}{\alpha} (\w_1\w_3)^{-x} -\left(\frac{\w_1}{\w_2}\right)^\frac{\alpha-d}{\alpha} (\w_1\w_2)^{-x}\right]\\
\nonumber & &\w_1^{\lambda+2-2x}\left[ \w_1^{2x-\lambda-2}-\w_2^{2x-\lambda-2}-\w_3^{2x-\lambda-2}\right]\,  \delta(\w_1\!-\!\w_2\!-\!\w_3)\, d\w_{23}\\
\label{eq-applyZT3}&=& \!\!\!\ckz^2\int\!\! K_1(\w_3,\w_2) (\w_1\w_2\w_3)^{-x}\,\w_1^\frac{\alpha-d}{\alpha} \left[\w_1^{x-\frac{\alpha-d}{\alpha}} -\w_2^{x-\frac{\alpha-d}{\alpha}} -\w_3^{x-\frac{\alpha-d}{\alpha}} \right]\\
\nonumber & &\w_1^{\lambda+2-2x}\left[ \w_1^{2x-\lambda-2}-\w_2^{2x-\lambda-2}-\w_3^{2x-\lambda-2}\right]\,  \delta(\w_1\!-\!\w_2\!-\!\w_3)\, d\w_{23}.
\end{eqnarray}
\end{widetext}
In these manipulations we have used Eqs.~(\ref{eq-defnK2K3}) and the fact that
in the integrand, $\w_2+\w_3 = \w_1$. It is now clear that the total
collision integral also vanishes when $x=\frac{\alpha-d}{\alpha} +1$ so that
\begin{equation}
\label{eq-thermodynamicSpectrum}
\N{\w} \sim \w^{-\left(\frac{\alpha-d}{\alpha}+1\right)}
\end{equation}
is also a stationary solution of Eq.~(\ref{eq-3WKEB}).
Using Eq.~(\ref{eq-transformToFreq}),  this spectrum 
 translates into $\n{k} \sim k^{-\alpha}$. The energy per mode
is then $\w_{\k} \n{k} = {\rm const}$. Hence Eq.~(\ref{eq-thermodynamicSpectrum})  corresponds to
the equilibrium solution. We note that the thermodynamic spectrum is one
aspect of Eq.~(\ref{eq-3WKE}) which cannot be expressed in terms of the 
parameter $\lambda$ introduced in Sec.~\ref{sec-3WKE}.

\subsection{Locality of the Kolmogorov-Zakharov Spectrum}

The Zakharov transformation used to obtain the stationary Kolmogorov-Zakharov
spectrum is only a valid procedure if the collision integral is convergent
on the KZ spectrum. This property should be checked a-posteriori and is
referred to as ``locality''. The choice of terminology comes from the
requirement that the collision integral in the inertial range should not
be dominated by the high or low frequency cut-offs. The determination
of locality is quite a delicate issue so in this section we shall perform
the analysis explicitly for Eq.~(\ref{eq-3WKEB}).

We can write Eq.~(\ref{eq-3WKEB}) in the form
\begin{eqnarray}
& &\nonumber \pd{\N{\w_1}}{t} =\int d\w_2d\w_3\left[R(\w_1,\w_2,\w_3)\,\delta^{\w_1}_{\w_2\,\w_3}\right.\\
& &\left.-R(\w_2,\w_3,\w_1)\,\delta^{\w_2}_{\w_3\,\w_1}-R(\w_3,\w_1,\w_2)\,\delta^{\w_3}_{\w_1\,\w_2} \right],
\end{eqnarray}
where
\begin{eqnarray}
\nonumber & &R(\w_1,\w_2,\w_3) = K_1(\w_3,\w_2)\N{\w_2}\N{\w_3}\\
\label{eq-defnR}& & -  K_2(\w_3,\w_2)\N{\w_1}\N{\w_3}  - K_3(\w_3,\w_2)\N{\w_2}\N{\w_2}.
\end{eqnarray}
Using Eqs.~(\ref{eq-defnK2K3}), this can be written
\begin{eqnarray}
\nonumber && R(\w_1,\w_2,\w_3)=K_1(\w_3,\w_2) \left[ \N{\w_2}\N{\w_3} \right.\\
& &\left. - {\scriptstyle \left(\frac{\w_2+\w_3}{\w_2}\right)^{\frac{\alpha-d}{\alpha}}}\N{\w_1}\N{\w_3} \label{eq-defnR2} - {\scriptstyle \left(\frac{\w_2+\w_3}{\w_3}\right)^{\frac{\alpha-d}{\alpha}}}\N{\w_1}\N{\w_2}\right].
\end{eqnarray}
Using the delta functions and integrating out $\w_3$, the collision integral
can then be written $S[\N{\w}] = \int d\w_2\,I(\w_1,\w_2)$ where
\begin{eqnarray}
\nonumber I(\w_1,\w_2) &=& Q(\w_2,\w_1-\w_2) - Q(\w_2-\w_1,\w_1)\\
\label{eq-I}& &  - Q(\w_1,\w_2),
\end{eqnarray}
and
\begin{equation}
\label{eq-defnQ}
Q(\w_i,\w_j) = R(\w_i+\w_j,\w_i,\w_j).
\end{equation}
We need to determine the convergence properties of Eq.~(\ref{eq-I}) as
$\w_2\to 0$ and $\w_2\to\infty$ for a general power law distribution,
$\N{\w}\sim \w^{-x}$ and then show that the integral is convergent when
$x$ is the K-Z value. This cannot be determined from simply counting
powers of $\w_2$ since there are some hidden cancellations which
occur. Indeed, if these cancellations did not occur, so that power counting
would work, it would be impossible to obtain a convergent integral since
no power of $\w_2$ can be integrable both at 0 and at $\infty$.

Let us first examine the limit $\w_2\to 0$. We need to determine the smallest
power of $\w_2$ in $I(\w_1,\w_2)$ as $\w_2\to 0$. In this case, the second term
in Eq~(\ref{eq-I}) vanishes and, after performing a Taylor expansion for 
small $\w_2$, we find
\begin{equation}
\label{eq-ILeftAsymptotics}
I(\w_1,\w_2) \stackrel{\scriptscriptstyle \w_2\to 0}{\sim} \w_2 \left.\pd{Q(y,\w_2)}{y}\right|_{y=\w_1} \hspace{-0.5cm} +\mbox{higher order in $\w_2$}
\end{equation}
We obtain an unexpected cancellation which makes the smallest power
of $\w_2$ larger by 1 than expected from simple power counting. Now we need
to determine the behaviour of $Q(\w_1,\w_2)$ as $\w_2\to 0$ when
$\N{\w}\sim \w^{-x}$. To accomplish this, we shall need to know the asymptotic
behaviour of $K_1(\w_i,\w_j)$. Let us introduce exponents $\mu$ and $\nu$
which characterise the asymptotics of $K_1(\w_i,\w_j)$ as follows:
\begin{equation}
\label{eq-muAndNuDefn}
K_1(\w_i,\w_j) \sim \w_i^\mu \w_j^\nu \hspace{0.5cm}\mbox{for $\w_1\gg\w_2$.}
\end{equation}
With some care we shall find that an additional cancellation occurs within the
structure of $Q(\w_1,\w_2)$. When $\N{\w} = \w^{-x}$, Eq.~(\ref{eq-defnQ}) and
Eq.~(\ref{eq-defnR2}) give:
\begin{widetext}
\begin{eqnarray}
\nonumber &&Q(\w_1,\w_2) = K_1(\w_2,\w_1)\left[\left(\w_1^{-x} - {\scriptstyle \left(\frac{\w_1+\w_2}{\w_1}\right)^{\frac{\alpha-d}{\alpha}}}(\w_1 + \w_2)^{-x}\right)\,\w_2^{-x} - {\scriptstyle \left(\frac{\w_1+\w_2}{\w_2}\right)^{\frac{\alpha-d}{\alpha}}}(\w_1 + \w_2)^{-x}\,\w_1^{-x} \right]\\
\nonumber && \sim \w_1^\mu\w_2^\nu\left[\left(- \w_2\left.\dd{}{y}{\scriptstyle \left(\frac{y}{\w_1}\right)^{\frac{\alpha-d}{\alpha}}}y^{-x}\right|_{y=\w_1} +O(\w_2^2)\right)\,\w_2^{-x} - {\scriptstyle \w_2^{\frac{d-\alpha}{\alpha}}}\,\w_1^{-2x} \right]\\
\label{eq-QLeftAsymptotics}&& = \w_1^\mu\w_2^\nu\, \left[ \w_2^{1-x}f(\w_1) +  \w_2^{\frac{d-\alpha}{\alpha}}g(\w_1)\right]
\end{eqnarray}
\end{widetext}
Two cases therefore arise depending on whether $1-x > \frac{d-\alpha}{\alpha}$
or $1-x < \frac{d-\alpha}{\alpha}$. From Eq.~(\ref{eq-thermodynamicSpectrum})
it is clear that these two cases correspond to the exponent, $x$, being
bigger or smaller respectively than the thermodynamic exponent. In the
former case, the small $\w_2$ behaviour of $Q(\w_1,\w_2)$ is $\w_2^{\nu-x+1}$.
Putting this together with Eq.~(\ref{eq-ILeftAsymptotics}), we see that the
small $\w_2$ behaviour of $I(\w_1,\w_2)$ is $\w_2^{\nu-x+2}$. In this
case, the condition for convergence of the collision integral at small $\w_2$
is $\nu-x+2 > -1$ which gives
\begin{equation}
\label{eq-localityAR}
x < \nu + 3.
\end{equation}
In the latter case, the small $\w_2$ behaviour of $Q(\w_1,\w_2)$ is $\w_2^{\nu+\frac{d-\alpha}{\alpha}}$.
The small $\w_2$ behaviour of $I(\w_1,\w_2)$ is then $\w_2^{\nu+\frac{d-\alpha}{\alpha}+1}$, in which case, the condition for convergence of the collision integral at small $\w_2$
is $\nu+\frac{d-\alpha}{\alpha}+1 > -1$ which gives
\begin{equation}
\label{eq-localityBR}
\frac{\alpha-d}{\alpha}+1 < \nu  +3.
\end{equation}

Let us now examine the limit $\w_2\to \infty$. We need to determine the largest
power of $\w_2$. For large, $\w_2$, there is no analogous cancellation between
the terms in Eq.~(\ref{eq-I}) which led us to Eq.~(\ref{eq-ILeftAsymptotics}).
In this case, the first term in Eq.~(\ref{eq-I}) vanishes and we find
\begin{equation}
\label{eq-IRightAsymptotics}
I(\w_1,\w_2) \stackrel{\scriptscriptstyle \w_2\to\infty}{\sim} -2\,Q(\w_1,\w_2).
\end{equation}
There is still, however, a cancellation {\em within} $Q(\w_1,\w_2)$. We perform
an analysis similar to that leading to Eq.~(\ref{eq-QLeftAsymptotics}), except
we Taylor expand in $\w_1$ since it is $\w_2$ which is large in this limit.
The result is:
\begin{equation}
\label{eq-QRightAsymptotics}
Q(\w_1,\w_2)\stackrel{\scriptscriptstyle \w_2\to\infty}{\sim} \w_1^\mu\w_2^\nu\, \left[ \w_2^{1-x}f(\w_1) +  \w_2^{\frac{d-\alpha}{\alpha}}g(\w_1)\right].
\end{equation}
There are again two casesm depending on whether $x$ is bigger or smaller than 
the thermodynamic exponent. In the
former case, the large $\w_2$ behaviour of $Q(\w_1,\w_2)$ is $\w_2^{\mu-x-1}$.
Putting this together with Eq.~(\ref{eq-IRightAsymptotics}), we see that the
large $\w_2$ behaviour of $I(\w_1,\w_2)$ is $\w_2^{\mu-x-1}$. In this
case, the condition for convergence of the collision integral at large $\w_2$
is $\mu-x-1 < -1$ which gives simply
\begin{equation}
\label{eq-localityAL}
x > \mu.
\end{equation}
In the latter case, the large $\w_2$ behaviour of $Q(\w_1,\w_2)$ 
and, hence $I(\w_1,\w_2)$, is $\w_2^{\mu+\frac{\alpha-d}{\alpha}-2x}$.
The condition for convergence of the collision integral at large $\w_2$
is then $\mu+\frac{\alpha-d}{\alpha}-2x < -1$ which gives
\begin{equation}
\label{eq-localityBL}
x > \frac{1}{2}\left(\mu+\frac{\alpha-d}{\alpha}+1\right).
\end{equation}

The conditions for convergence of the collision integral are different
depending on whether the exponent, $x$, is greater than or less than
the thermodynamic exponent given by Eq.~(\ref{eq-thermodynamicSpectrum}).
In the former case, we must satisfy Eq.~(\ref{eq-localityAL}) and
Eq.~(\ref{eq-localityAR}). Putting these two together, we see that a 
range of $x$ exists which produce a convergent collision integral if
\begin{equation}
\label{eq-localityCriterionMuNu}
\mu < \nu+3.
\end{equation}
We note that when such a range exists, the K-Z exponent is at the
centre of this range and is therefore local. In the latter case, we must satisfy
Eq.~(\ref{eq-localityBL}) and Eq.~(\ref{eq-localityBR}) which together can 
be rearranged to give
\begin{equation}
x > \frac{1}{2}\left(\mu+\nu + 3\right) = x_{\rm KZ}.
\end{equation}
Hence we conclude that the K-Z spectrum cannot be local if it is 
shallower than the thermodynamic spectrum. Hence the conditions for locality
of the K-Z spectrum are
\begin{eqnarray}
\label{eq-localityCriterion}
x_{\rm KZ} &>& 1+\frac{\alpha-d}{\alpha}\\ 
\nonumber \mu &<& \nu+3.
\end{eqnarray}
Note that, provided the K-Z spectrum is steeper than the thermodynamic
one, the condition for locality only depends on the properties of 
$K_1(\w_i,\w_j)$ and not on $\alpha$ and $d$.

\subsection{Physical Examples}
To close this very brief review of wave turbulence, let us calculate the
value of $\lambda$ from Eq.~(\ref{eq-lambdaDefn}) for some of the commonest 
physical examples of 3-wave
turbulence (see \cite{CNN03} for a summary). Capillary waves on deep water, 
the archetypal example of 3-wave turbulence, have $\alpha=\frac{3}{2}$, $d=2$
and $\beta=\frac{9}{4}$ giving $\lambda=2$. Acoustic turbulence has $\alpha=1$,
$d=3$ and $\beta=\frac{3}{2}$ giving again $\lambda=2$. Quasi-2D Alfv\'{e}n
wave turbulence has $\alpha=1$, $d=2$ and $\beta=2$ which yields
$\lambda=3$.

\section{Choices of Spectral Truncation}
\label{sec-truncation}
It is of interest to study Eq.~(\ref{eq-3WKE}) in the presence of a 
frequency cut-off, which we shall denote by $\W$. In the truncated system,
all $\N{\w}$ having $\w>\W$ are taken to be zero. In the discrete case, this 
corresponds to studying a truncated version of the infinite set of ODEs 
constituting the kinetic equation. This interest may be forced upon us. In
a numerical setting, for example, we must necessarily discretise and truncate in
order to develop a computational scheme. Numerical practicalities aside,
spectral truncations of turbulent systems have recently been of considerable 
theoretical interest in their own right because of the connection between
spectral trunction and the phenomenon of thermalisation. Thermalisation refers
to a situation in which a turbulent system exhibits a mixture of constant flux 
and equipartition behaviour. We shall have more to say about thermalisation
in Sec.~\ref{sec-thermalisation} but first, let us clarify some issues
related to the implementation of the spectral cut-off.

We shall truncate the discrete kinetic equation, Eq.~(\ref{eq-discreteKE}).
Requiring that $N_i=0$ for $i>\W$ does not uniquely determine the
resulting set of equations. We must choose what to do with loss terms in the
forward transfer integral, Eq.~(\ref{eq-discreteKE}), for which $i+j > \W$.
In the graphical representation of Fig.~\ref{fig-S}, the issue is how to
treat the set of triads for which $N_j$ and $N_k$ are in the truncated
system but $N_i$ is not. There is no ambiguity arising from such triads in
the backscatter terms, Eq.~(\ref{eq-discreteS2}) and Eq.~(\ref{eq-discreteS3}). From
the rates, Eq.~(\ref{eq-S2Rates}) and Eq.~(\ref{eq-S3Rates}), that if 
$\w_i>\W$, then $N_i=0$ and all resulting rates are zero. There is, however, an 
ambiguity arising from the forward transfer term, Eq.~(\ref{eq-discreteS1}).
From the energy transfer rates, Eq.~(\ref{eq-S1Rates}), it is clear that 
having $\w_i>\W$ and $N_i=0$ does not necessarily imply that the rates are
zero although terms having $\w_i>\W$ can only {\em decrease} the total number
of waves. They correspond to transfer of energy across the cutoff from the
interaction of two waves which are themselves, below the cutoff. Such 
interactions should therefore be treated as dissipation terms in the
truncated system. Let us separate these terms from the rest and write
the truncated system as follows, ignoring the external forcing and
dissipation terms for now:

\begin{equation}
\label{eq-truncatedKE}\dd{N_i}{t} = S_i[{\bf N},\W] -  \gamma\,T_i[{\bf N},\W]\ \ \ \ \mbox{$i=1,2,\ldots\W$}
\end{equation}
where
\begin{equation}
\label{eq-truncatedS}
S_i[{\bf N},\W]  = \,S_i^{(1)}[{\bf N},\W] +  \,S_i^{(2)}[{\bf N},\W] +  \,S_i^{(3)}[{\bf N},\W]
\end{equation}
with the truncated collision integrals given by
\begin{eqnarray}
\nonumber S_i^{(1)}[{\bf N},\W] =& & \sum_{j=1}^{i-1} K_1(j,i-j)\, N_j N_{i-j}\\
\label{eq-truncatedS1} &-& \sum_{j=i+1}^{\W} K_1(j-i,i)\, N_i N_{j-i}\\
\nonumber  &-&  \sum_{j=1}^{\W-i} K_1(i,j)\,N_i N_{j},
\end{eqnarray}
\begin{eqnarray}
\nonumber S_i^{(2)}[{\bf N},\W] = &-& \sum_{j=1}^{i-1} K_2(j,i-j)\, N_i N_{j}\\
\label{eq-truncatedS2} &+& \sum_{j=i+1}^{\W} K_2(j-i,i)\, N_j N_{j-i} \\
\nonumber  &+& \sum_{j=1}^{\W-i} K_2(i,j)\,N_i N_{i+j}
\end{eqnarray}
and
\begin{eqnarray}
\nonumber S_i^{(3)}[{\bf N},\W] = &-&  \sum_{j=1}^{i-1} K_3(j,i-j)\, N_i N_{i-j}\\
\label{eq-truncatedS3} &+& \sum_{j=i+1}^{\W} K_3(j-i,i)\, N_j N_{i} \\
\nonumber  &+& \sum_{j=1}^{\W-i} K_3(i,j)\,N_j N_{i+j}
\end{eqnarray}
The dissipation terms discussed above corresponding to transfer of energy
across the cutoff have been gathered together into 
\begin{eqnarray}
\label{eq-T}
T_i[{\bf N},\W] &=& \gamma\,\left(\sum_{j=\W+1}^{\W+i} K_1(j-i,i)\,N_i N_{j-i}\right.\\
\nonumber &&\left.+\sum_{j=\W-i+1}^{\W} K_1(i,j)\,N_i N_{j}\right),
\end{eqnarray}
We can now select between different natural truncations by varying the 
parameter $\gamma$ in Eq.~(\ref{eq-truncatedKE}). Taking $\gamma=0$ 
corresponds to discarding triads which transfer energy across the
cutoff. We shall refer to this as the {\em closed} truncation. For the
closed truncation, the energy, $E_\W$, of the truncated system, 
\begin{equation}
E_\W = \sum_{i=1}^\W i\,N_i,
\end{equation}
is conserved by Eq.~(\ref{eq-truncatedKE}). Taking $\gamma=1$ means that
we allow energy to freely cross the cutoff at which point it is removed from the
system (dissipated). We shall refer to this as the {\em open} truncation.
With the open truncation, the energy, $E_\W$, of the truncated system
may decrease as a function of time. Taking $0 < \gamma< 1$  corresponds to
something intermediate between the open and closed truncations which we
shall refer to as a {\em partially open} truncation. 

There is nothing intrinsic to the system which tells us which truncation we
should choose. It will depend on what we want to do. It is often the
case that we consider the truncated system as being an approximation of the
original kinetic equation and would like to recover the original dynamics
when we take $\W\to \infty$. As we shall see in 
Sec.~\ref{sec-dissipativeAnomaly}, the choice of truncation is sometimes
irrelevant for recovering the original dynamics as $\W\to \infty$ and sometimes
essential.

\section{A New Numerical Algorithm for Solving the 3-Wave Kinetic Equation}
\label{sec-numerics}
Recasting the 3-wave kinetic equation, Eq.~(\ref{eq-3WKE}) as an aggregation--
fragmentation problem has another advantage in addition to the conceptual
clarity discussed in Sec.~\ref{sec-WTResults}. It is the basis of a new 
numerical method for solving Eq.~(\ref{eq-3WKE}) accurately over a large range 
of frequency scales. This method is an extension to Eq.~(\ref{eq-3WKEB}) of an 
elegant method developed by M.H. Lee \cite{LEE2000,LEE2001} to solve the 
Smoluchowski equation, Eq.~(\ref{eq-SKE}). In this section we shall give the
details of this method.

Before proceeding, we would like to remark that the objective is to design a
numerical method which can solve the isotropic kinetic equation, 
Eq.~(\ref{eq-3WKEB}), accurately over many decades of frequency for an 
arbitrary interaction coefficient with the objective of studying the scaling 
properties of this idealised system. This is in contrast with, but 
complementary to, the majority of numerical effort invested in solving
wave turbulence kinetic equations which has focused on approximating 
the 4-wave kinetic equation for the specific case of deep water gravity
waves under anisotropic conditions \cite{HHAB1985,ZP1999,PF2002} which is
strongly motivated by the applications to wave forecasting.

Any numerical scheme requires that we work with a  set of $N$ discrete 
frequencies, $\w_i=\w_0+(i-1)\Delta\w$, where $i=1,\ldots N$, $\Delta\w$ is 
the mode spacing and $\w_{N}=\W$ is the frequency cut-off.  We shall therefore 
take $\w_0=1$ and $\Delta\w=1$ and work with Eq.~(\ref{eq-discreteKE}). 
It is more  natural to work with the energies of the modes, rather than their 
corresponding occupation numbers, ${\bf N}$.  ${\bf E}(t)$, the vector of 
energies is obtained from ${\bf N}(t)$ by the relation 
$E_i=i\,N_i$, $i=1,\ldots,\W$.

The objective is to solve the following set of coupled nonlinear 
ordinary differential equations for the $E_i$ obtained by multiplying
Eq.~(\ref{eq-discreteKE}) by $i$:
\begin{eqnarray}
\label{eq-discreteKE2}\dd{E_i}{t} &=& i\,S_i[{\bf N},\W]  -\gamma\,i\,T[{\bf N},\W]\\
\nonumber & & +  i\,F_i[{\bf N}] -  i\,D_i[{\bf N}]
\end{eqnarray}
with $S_i[{\bf N},\W]$ and $T[{\bf N},\W]$ given by Eq.~(\ref{eq-truncatedS})
and Eq.~(\ref{eq-T}) respectively. Depending on the application, it might
also be useful to keep track of the cumulative energy, $E_D(t)$, dissipated by
the $T[{\bf N},\W]$ terms. In such cases, we supplement 
Eq.~(\ref{eq-discreteKE2}) with the following equation:
\begin{equation}
\nonumber \dd{E_D}{t} = \gamma\,\sum_{i=1}^\W\,i\,T_i[{\bf E}].
\end{equation}
We shall take the forcing to be
\begin{equation}
F[{\bf E}] = J\, \delta_{i,1}
\end{equation}
so that the total rate of injection of energy into the system is $J$. We
take $D_i[{\bf N}]$ to be zero since dissipation is provided by $T[{\bf N},\W]$.
Then setting $\gamma=1$ we recover the open truncation. Setting
$\gamma=0$ we recover the closed truncation and by setting $\gamma$ somewhere
between, we get a partially open truncation.

There are several problems to be overcome in solving Eqs.~(\ref{eq-discreteKE2}).
Firstly  Eqs.~(\ref{eq-discreteKE2}) become very stiff as $\lambda$ increases.
This means that when straightforward explicit integration routines are
applied to the system, the time step required to maintain numerical 
stability remains very small even when the solution we are trying to compute
is varying slowly. The result is that it takes an impractically long time to
compute the solution with explicit methods and one must resort to an
implicit integration algorithm in many cases. The second problem is that
we are interested in situations when the number of modes involved
is large but the number of modes one can deal with by direct integration of
 Eqs.~(\ref{eq-discreteKE2}) is limited by the computational cost of evaluating
the sums on the righthand side. This issue is addressed by the technique
developed in \cite{LEE2000}. The modes are grouped  into exponentially spaced 
bins and the net exchange of energy between triads of bins is approximated
rather than the exchange of energy between triads of individual modes.

\subsection{Time-stepping}

\begin{figure}
\includegraphics[width=6.5cm]{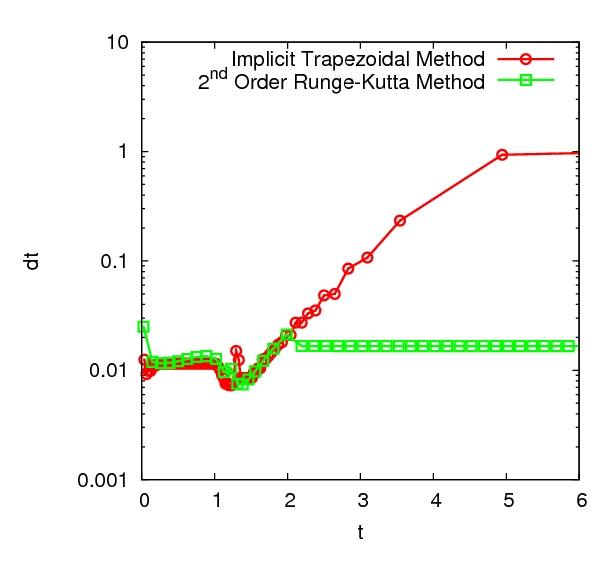}\\
\caption{\label{fig-stiffness} Illustration of the stiffness of 
Eq.~(\ref{eq-discreteKE2}) with the $\lambda=2 $ sum kernel, 
Eq.~(\ref{eq-sumKernel}). $\W$ was taken to be 100, the energy injection rate
was $J=1$ and the open truncation was used. The plot compares the timesteps 
required to keep the numerical solution within an error tolerance of $1\times
10^{-6}$ using the RK2 algorithm and IT method.}
\end{figure}

The solutions of Eq.~(\ref{eq-discreteKE2}) exhibit scaling behaviour, with 
the result that there is a wide 
variation of the timescale during the course of the evolution. This 
necessitates \cite{LEE2000} the use of adaptive timestepping to keep the error 
within a prescribed limit \cite{NUMREC}. The explicit second order Runge-Kutta 
(RK2) method and implicit trapezoidal (IT) method were both used in conjunction with
a step-doubling procedure to adjust the timestep, $h$. Both have stepwise
errors of $O(h^3)$. The use of an implicit method is necessary for larger
values of $\lambda$ because Eq.~(\ref{eq-discreteKE2}) becomes increasing stiff for 
larger values of $\lambda$.  This is illustrated clearly in 
Fig.~\ref{fig-stiffness} which compares the stepsize required by the explicit
RK2 and implicit IT integration algorithms in order to maintain a given
error as the solution of Eq.~(\ref{eq-discreteKE2}) with the sum kernel, 
Eq.~(\ref{eq-sumKernel}) with $\lambda=2$. Initially, the dynamics is fast
an both methods require small timesteps to keep the error within the 
prescribed tolerance. Around $t=2$ the solution approaches the stationary
state and the dynamics slows down. The explicit RK2 algorithm continues to
require small steps despite the fact that the solution is no longer evolving
quickly. In contrast the implicit IT algorithm can take increasingly large 
steps as the dynamics slows down.
This behaviour is the classical symptom of a stiff system and renders
explicit solvers practically useless for solving the 3WKE with larger
values of $\lambda$.

There is a price to be paid for dealing with the stiffness issue. The IT method requires that we solve the following set
of implicit nonlinear equations (we ignore the forcing and dissipation terms
for now since they are straightforward to include) to find the energies at the next timestep, 
$E_i(t+h)$, from the energies at the current timestep, $E_i(t)$:
\begin{equation}
E_i(t+h) = E_i(t)  + \frac{h}{2}\, \left( S_i[{\bf E}(t)] +  S_i[{\bf E}(t+h)]\right)
\end{equation}
This was done using the GSL implementation \cite{GSL} of the Rosenbrock 
algorithm \cite{ROS1960}, a standard method of
multi-dimensional root finding. The current values, $E_i(t)$, were used
as the initial guess for the root finding procedure.

\subsection{Coarse-graining}

A direct integration of Eq.~(\ref{eq-discreteKE}) is practical only for 
relatively small numbers of modes. In order to resolve large inertial ranges,
we coarse-grain the modes into bins and approximately compute the net energy
transfer rate between bins following the approach developed in 
\cite{LEE2000} for the Smoluchowski equation. 

We need to divide the frequency domain, $[1,\W)]$, up into $N$ bins. We shall
adopt the notation
\begin{displaymath}
B_i = [\w_i^L, \w_i^R)
\end{displaymath}
to denote the $i^{\rm th}$ bin and denote the bin widths by
\begin{displaymath}
\Delta\w_i =  \w_i^R -  \w_i^L.
\end{displaymath}
We use the same bin structure as adopted in 
\cite{LEE2000}: the first $n$
bins are linearly spaced and the next $N-n$ bins are defined by a 
geometric sequence of boundary points having ratio $a=10^\frac{1}{n}$. The
final bin is defined so that its right boundary is $\W$. With this definition,
there are approximately $n$ bins per decade of frequency space with the
total number of bins, $N$, determined by the value of the frequency
cut-off, $\W$. We define a characteristic frequency, $\W_i$, of each bin by
\begin{displaymath}
\W_i = \frac{1}{2}(\w_i^L + \w_i^R).
\end{displaymath}
We shall continue to use $E_i$ to denote the total amount of energy contained
in bin $i$ despite the fact that for $i>n$ this now refers to an entire 
bin of frequencies rather than an individual frequency (for the first $n$
bins, each of which contains only a single mode, $E_i$ remains just the
energy contained in that mode). Likewise we shall continue to use $N_i$
to denote the total number of waves in bin $i$.  Within each bin, $B_i$,
with $i>n$ the energy  and wave-action distributions within the bin are 
approximated with power law distributions:
\begin{eqnarray}
e_i(\w) &=& a_i\left(\frac{\w}{\w_i^L}\right)^{b_i}\hspace{1.0cm} \w_i^L \leq \w < \w_i^R\\
n_i(\w) &=& a_i\left(\frac{\w}{\w_i^L}\right)^{b_i-1}.
\end{eqnarray}
The exponent, $b_i$, is obtained by interpolating the characteristic  energies 
of the neighbouring bins:
\begin{equation}
a_i = \frac{\log\left(\frac{E_{i+1}}{\Delta\w_{i+1}}\right) - \log\left(\frac{E_{i-1}}{\Delta\w_{i-1}}\right)}{\log \W_{i+1} - \log \W_{i-1}},
\end{equation}
and the prefactor, $a_i$, is fixed by the normalisation
\begin{equation}
E_i = \int_{\w_i^L}^{\w_i^R}d\w\,e_i(\w).
\end{equation}

We have seen in Sec.~\ref{sec-3WKE} that computing the collision integral on 
the RHS of Eq.~(\ref{eq-discreteKE}) is equivalent to computing the appropriate
rates of energy exchange given by Eq.~(\ref{eq-S1Rates}), Eq.~(\ref{eq-S2Rates})
and Eq.~(\ref{eq-S3Rates}) for the members of each resonant triad and 
summing over all triads. After course-graining, we need to compute
the net rates of energy transfer resulting from the interactions of all
triads $(\w_i, \w_j, \w_k)$ for which $\w_i \in B_i$ and $\w_j \in B_j$ 
(of course we must allow the possibility that $B_i$=$B_j$) and sum the
results over all possible pairs of bins $B_i$ and $B_j$. The fundamental
quantity which we evolve is naturally the total amount of energy contained
in each bin.

When considering interactions between a pair of bins, $B_j$ and $B_k$, we 
shall adopt the labeling used in Fig.~\ref{fig-S}: bin $j$ to the left of bin 
$k$.  In computing the rates of transfer of energy generated by these 
interactions the first step is to determine which bin or bins contain the 
modes in resonance with those in $B_j$ and $B_k$. 
Before describing how the subsequent energy redistribution works in detail, let
us explain the key approximation which will used.
Consider, for example, the 
forward-scatter process. The total rate of energy transfer to higher 
frequencies due to resonances between modes in $B_i$ and $B_k$ is given
by a double integral,
\begin{eqnarray}
\nonumber \Delta E &=& \int_{\w_j^L}^{\w_j^R} d\w_j \int_{\w_k^L}^{\w_k^R} d\w_k \, (\w_j + \w_k)\,K_1(\w_j,\w_k)\\
& &\hspace{2.0cm}\,n_j(\w_j)\,n_k(\w_k),
\end{eqnarray}
and these rates of gain of energy are distributed (non-uniformly) among a set 
of modes with frequencies lying between $\w_j^L+\w_k^L$ and $\w_j^R+\w_k^R$.
The key approximation which we make is to treat all modes in $B_j$ (the lower
frequency narrower bin) as having frequency $\W_j$ (of course, this is
not even an approximation if $j<=n$). That is
we replace $n_j(\w_j)$ with $N_j\,\delta(\w_j-\W_j)$ so that instead of
computing double integrals, we need to compute one dimensional integrals of 
the form
\begin{equation}
\Delta E = N_j \int_{\w_k^L}^{\w_k^R} d\w_k \, (\W_j + \w_k)\,K_1(\W_j,\w_k)\,n_k(\w_k)
\end{equation}
with the gains of energy distributed  among a set of modes with 
frequencies lying between $\W_j+\w_k^L$ and $\W_j+\w_k^R$.

With this approximation in mind, let us now describe explicitly the
calculation of the rates of energy transfer. For each pair of 
bins, $B_j$ and $B_k$, there are three possibilities which must be
treated separately. They are, as listed below, labeled $A$, $B$ and $C$.
With apologies for the somewhat clumsy notation, we now list the energy
transfer rates to/from each bin for each of these cases split up
according to the contributions from each of the processes described by
$S_1[\N{\w}]$, $S_2[\N{\w}]$ and $S_3[\N{\w}]$. $L$ shall denote an energy 
loss term and $G$ an energy gain term. The actual integrals are written 
explicitly in appendix \ref{sec-app2} to avoid an overload of technical
details.
\begin{itemize}
\item{{\bf Case A} : $k\leq n$\\}
This is the simplest case in which $B_k$ (and hence $B_j$) is a discrete
bin. Given $j$ and $k$, we identify the bin, $B_i$, which contains 
$\w_i=\W_j+\W_k$. Whether it is a discrete or continuous bin is irrelevant.
The transfer rates can be read off almost immediately from 
Eqs.~(\ref{eq-S1Rates}), Eqs.~(\ref{eq-S2Rates}) and Eqs.~(\ref{eq-S3Rates}):
\begin{eqnarray}
\nonumber (\Delta E)_j &=& -^{A}L_j^{(1)} + ^AG_j^{(2)} + ^AG_j^{(3)}\\
\label{eq-CaseA}(\Delta E)_k &=& -^AL_k^{(1)} + ^AG_k^{(2)} + ^AG_k^{(3)}\\
\nonumber (\Delta E)_i &=& (\Delta E)_j  + (\Delta E)_k
\end{eqnarray}
where
\begin{eqnarray*}
^{A}L_j^{(1)} &=& \W_j\,K_1(\W_j,\W_k)\,N_j\,N_k\\
^{A}L_k^{(1)} &=& \W_k\,K_1(\W_j,\W_k)\,N_j\,N_k\\
^{A}G_j^{(2)} &=& \W_j\,K_2(\W_j,\W_k)\,N_j\,\frac{N_i}{\Delta\w_i}\\
^{A}G_k^{(2)} &=& \W_k\,K_2(\W_j,\W_k)\,N_j\,\frac{N_i}{\Delta\w_i}\\
^{A}G_j^{(3)} &=& \W_j\,K_3(\W_j,\W_k)\,N_k\,\frac{N_i}{\Delta\w_i}\\
^{A}G_k^{(3)} &=& \W_k\,K_3(\W_j,\W_k)\,N_k\,\frac{N_i}{\Delta\w_i}.
\end{eqnarray*}

\item{{\bf Case B} : $k>n$ and $[\W_j+\w_k^L, \W_j+\w_k^R]$ is contained in a single bin, $B_i$\\}
In this case, we identify the bin, $B_i$, which contains all modes in the
range $[\W_j+\w_k^L, \W_j+\w_k^R]$. The rates of energy transfer are
\begin{eqnarray}
\nonumber (\Delta E)_j &=& -^BL_j^{(1)} + ^BG_j^{(2)} + ^BG_j^{(3)}\\
\label{eq-CaseB}(\Delta E)_k &=& -^BL_k^{(1)} + ^BG_k^{(2)} + ^BG_k^{(3)}\\
\nonumber (\Delta E)_i &=& (\Delta E)_j  + (\Delta E)_k
\end{eqnarray}
with the relevant integrals provided in the appendix.

\item{{\bf Case C} : $k>n$ and $[\W_j+\w_k^L, \W_j+\w_k^R]$ is split between two bins, $B_{i_L}$ and $B_{i_R}$\\}
It can happen that the range of modes $[\W_j+\w_k^L, \W_j+\w_k^R]$ spans
two bins. With the above definition of the bin structure it is never more
than two. We identify these bins as $B_{i_L}$ and $B_{i_R}$. We must also
identify the point $\W^*=\w_{i_R}^L - \W_j \in B_k$ which marks the boundary
between those modes in $B_k$ which are resonant with modes in $B_{i_L}$ and
those modes in $B_k$ which are resonant with modes in $B_{i_R}$. Energy 
transfer is then split appropriately between $B_{i_L}$ and $B_{i_R}$:
\begin{eqnarray}
\nonumber (\Delta E)_j &=& _L(\Delta E)_j + _R(\Delta E)_j\\
\label{eq-CaseC}(\Delta E)_k &=& _L(\Delta E)_k + _R(\Delta E)_k\\
\nonumber (\Delta E)_{i_L} &=&  _L(\Delta E)_j +  _L(\Delta E)_k\\
\nonumber (\Delta E)_{i_R} &=&  _R(\Delta E)_j +  _R(\Delta E)_k
\end{eqnarray}
where
\begin{eqnarray}
\nonumber _L(\Delta E)_j &=& -^C_LL_j^{(1)} + ^C_LG_j^{(2)} + ^C_LG_j^{(3)}\\
\nonumber _R(\Delta E)_j&=& -^C_RL_j^{(1)} + ^C_RG_j^{(2)} + ^C_RG_j^{(3)}\\
\label{eq-CaseCDefns}_L(\Delta E)_k &=& -^C_LL_k^{(1)} + ^C_LG_k^{(2)} + ^C_LG_k^{(3)}\\
\nonumber  _R(\Delta E)_k&=& -^C_RL_k^{(1)} + ^C_RG_k^{(2)} + ^C_RG_k^{(3)}.
\end{eqnarray}
Again, the actual integrals are written explicitly in the appendix.
\end{itemize}

\begin{figure}
\includegraphics[width=6.5cm]{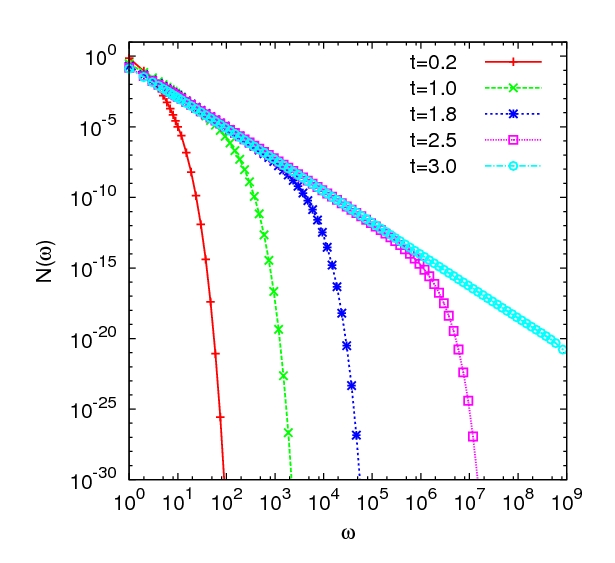}\\
\caption{\label{fig-exampleResults} Sample output from implementation the 
algorithm described in Sec.~\ref{sec-numerics}. This is a decay simulation
with $\lambda=3/2$ product kernel, Eq.~(\ref{eq-productKernel}).}
\end{figure}

Some typical outputs resulting from the implementation of the algorithm 
described in this section is shown in Fig.~\ref{fig-exampleResults}.

\section{Numerical Validation: Calculation of the Kolmogorov-Zakharov Constant}
\label{sec-validation}
The evolution of the energy spectrum shown in 
Fig.~\ref{fig-exampleResults} certainly looks plausible. Nevertheless, the approximations
made in deriving the algorithm described in Sec.~\ref{sec-numerics} are
not systematic and no convergence results have been proven. Therefore it
is essential to validate the code. From the discussions of 
Sec.~\ref{sec-3WKE}, it is obviously inadequate to use the measurement of
scaling exponents as a means of validation since the scaling properties of
Eq.~(\ref{eq-3WKEB}) and Eq.~(\ref{eq-SKE}) are practically identical even
though the physics is very different for the two equations. For Lee's
original implementation of this method for Smoluchowski equation, one had the 
luxury of several exact solutions \cite{LEE2000,LEY2003} against which
the full time-dependent evolution could be validated. There are, to the best
of our knowledge, no known exact solutions of the 3-wave kinetic equation
whch could play a similar diagnostic role in the present context. We
suggest instead, to use the measurement of the Kolmogorov-Zakharov constant,
$\ckz$,
as a diagnostic. It can be computed exactly as we shall now show. Furthermore 
its value is dependent on getting the internal structure of the collision
integral correct. Taken together with the measurement of stationary scaling
exponents, the measurement of $\ckz$ is a stringent test which the code was
required to pass.

Let us now calculate $\ckz$. Consider the statement of conservation of energy,
\begin{equation}
\label{eq-energyConservation}
\partial_{t}\left(\w_1\N{\w_1}\right) = \w_1\,S[\N{\w}] = -\partial_{\w_1} J_{\w_1},
\end{equation}
where $J_{\w}$ is the energy flux at frequency $\w$.
For a general power law spectrum, $\N{\w} = c\,\w^{-x}$, we can write the
right hand side like Eq.~(\ref{eq-applyZT1}). Introducing
rescaled integration variables, $u$ and $v$ defined by
\begin{displaymath}
\w_2 = \w_1\,v\ \ \ \ \w_3 = \w_1\,u,
\end{displaymath}
and integrating out $v$ allows us to write 
Eq.~(\ref{eq-energyConservation}) as
\begin{equation}
\partial_{\w_1} J_{\w_1} = - c^2\,\w_1^{\lambda-2\x+2}\,I(x),
\end{equation}
where
\begin{eqnarray}
& &I(x) = \int_0^1 \left[ K_1(u,1-u)\, (u(1-u))^{-x} \right.\\
\nonumber &&\left. -K_2(u,1-u)\, u^{-x} -K_3(u,1-u)\, (1-u)^{-x}\right]\\
\nonumber && \left[ 1-(1-u)^{2x-\lambda-2}-u^{2x-\lambda-2}\right]\,  du.
\end{eqnarray}
Integrating once, we get
\begin{equation}
J_{\w} = -\w^{\lambda-2 x +3} \frac{c^2 I(x)}{\lambda-2 x +3}.
\end{equation}
The energy flux, $J_{\w}$, should be constant, independent of $\w$, and equal
to $J$, the rate of energy injection, when $x$  takes the KZ value, 
$(\lambda+3)/2$. The limit needs to be taken using l'H\^{o}pital's
Rule since $I(x_{\rm KZ})=0$. The result is
\begin{equation}
J = \frac{1}{2}\, c^2\,\left.\dd{I}{x}\right|_{x=x_{\rm KZ}}.
\end{equation}
The K-Z constant, is therefore given by
\begin{equation}
\label{eq-KZconstant}
c_{\rm KZ} = \sqrt{2\,J\,\left.\dd{I}{x}\right|_{x=x_{\rm KZ}}^{-1}}
\end{equation}
The integral $\left.\dd{I}{x}\right|_{x=x_{\rm KZ}}$, can be calculated 
numerically, and analytically for some interaction coefficients. We shall
use this result to validate the numerical solution procedure.

\begin{figure}
\includegraphics[width=6.5cm]{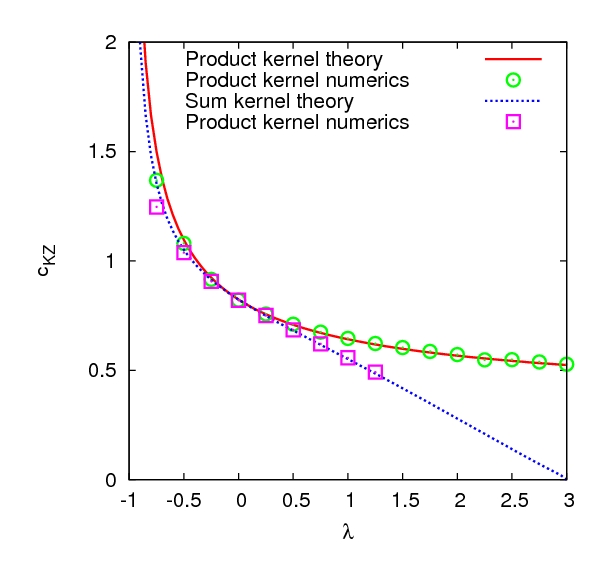}\\
\caption{\label{fig-validation} Numerical values of the Kolmogorov--Zakharov 
constant for the sum kernel, Eq.~(\ref{eq-sumKernel}) and the product kernel, 
Eq.~(\ref{eq-productKernel}), for a  range of values of $\lambda$. The 
theoretical curves come from numerical evaluation of Eq.~(\ref{eq-KZconstant}).
}
\end{figure}

Based on the discussion of the previous section,
very few properties of the kinetic equation actually depend on $\alpha$ or $d$
and those that do are all related to thermodynamic equilibrium. We shall 
therefore take $\alpha=d$ so that $K_1(\w_i,\w_j) = K_2(\w_i,\w_j) =
K_3(\w_i,\w_j)$. For such systems, the thermodynamic equilibrium spectrum is 
$\w^{-1}$.  We tested our code with both the product kernel, 
Eq.~(\ref{eq-productKernel}), and the sum kernel, Eq.~(\ref{eq-sumKernel}),
for various values of $\lambda$. This was done by allowing the system
to reach a stationary state with a constant rate of injection and then
compensating the computed spectrum, $\N{\w}$, with the K-Z spectrum and
fitting a constant to the result. The results are shown in 
Fig.~\ref{fig-validation}. The numerically fitted values of $c_{\rm KZ}$
are in excellent agreement with the theoretical predictions obtained from
Eq.~(\ref{eq-KZconstant}) for both kernels which gives us confidence in
the numerical algorithm introduced in Sec.~\ref{sec-numerics}. Note that the
value of the $c_{\rm KZ}$ diverges at $\lambda=-1$ for both kernels. This 
comes from the fact that, in the numerics we took $\alpha=d$ so the the
thermodynamic spectrum, Eq.~(\ref{eq-thermodynamicSpectrum}) has exponent 
$x=1$. Thus
at $\lambda=-1$ the thermodynamic and K-Z exponents coincide. From the locality
conditions, Eq.~(\ref{eq-localityCriterion}), derived in 
Sec.~\ref{sec-WTResults}, we know that the collision integral
diverges when this occurs due to the violation of the first 
condition in Eq.~(\ref{eq-localityCriterion}). Notice also that the vanishing of $c_{\rm KZ}$ 
for the sum kernel at $\lambda=3$ corresponds to the breakdown of
locality due to the violation of the second condition in 
Eq.~(\ref{eq-localityCriterion}). Fig.~\ref{fig-validation} therefore provides a nice consistency 
check on our earlier theoretical analysis.

\section{Finite and Infinite Capacity Cascades and Dissipative Anomaly}
\label{sec-dissipativeAnomaly}
The concept of a dissipative anomaly is key to understanding the statistical
properties of many turbulent systems. It refers to a situation in which the
average rate of energy dissipation tends to become independent of the
dissipation parameter in the dynamical equations in the limit where this 
dissipation parameter is  taken to zero. This phenomenon is explicitly 
demonstrable in Burgers' equation (see \cite{BK2007} for a review) where the 
rate of dissipation of energy in shocks becomes independent of the viscosity,
$\nu$, as $\nu \to 0$. It is also believed to be very relevant 
(see the discussion in \cite{EY2008}) for the high Reynolds number limit of 
hydrodynamic turbulence. 

\begin{figure}
\includegraphics[width=6.5cm]{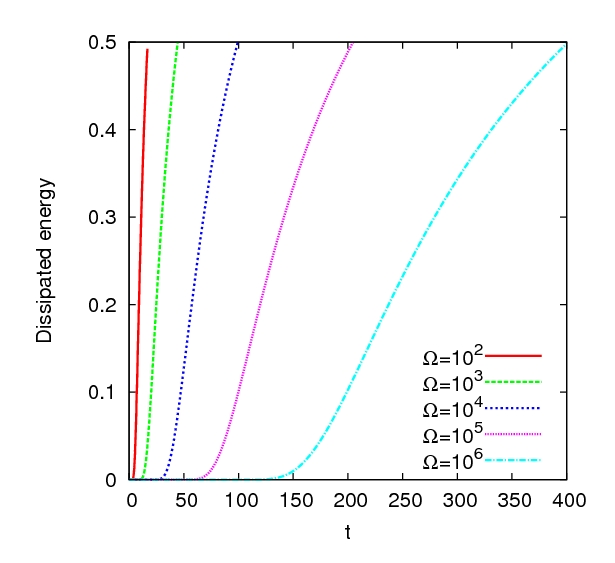}\\
\caption{\label{fig-infiniteCapacity} Time dependence of the dissipated
energy for increasing values of the dissipation scale, $\W$, for the product
kernel, Eq.~(\ref{eq-productKernel}) with $\lambda=\frac{3}{4}$.}
\end{figure}

\begin{figure}
\includegraphics[width=6.5cm]{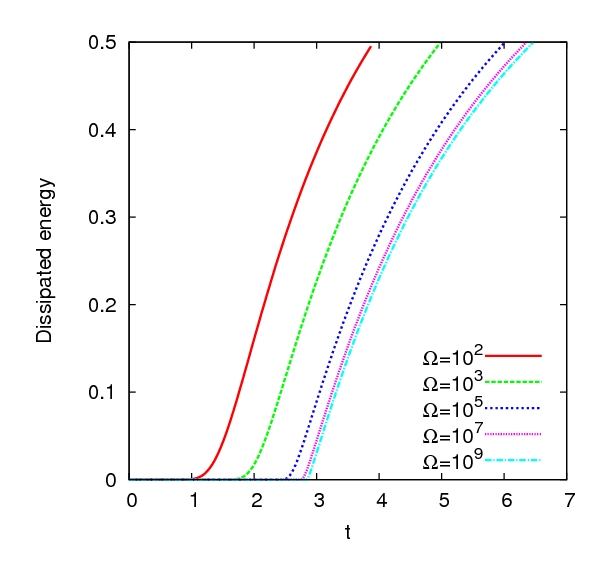}\\
\caption{\label{fig-finiteCapacity} Time dependence of the dissipated
energy for increasing values of the dissipation scale, $\W$, for the product
kernel, Eq.~(\ref{eq-productKernel}) with $\lambda=\frac{3}{2}$.}
\end{figure}

In wave turbulence it is expected that the dissipative anomaly is present for
finite capacity wave systems and absent for infinite capacity systems. In
Sec.~\ref{sec-WTResults} we showed that $\lambda=1$ is the boundary
between infinite and finite capacity. The difference between the two can be 
explicitly demonstrated using the numerical scheme developed in 
Sec.~\ref{sec-numerics}.  We have seen that an open truncation of the 3 wave
kinetic equation introduces dissipation of energy with the dissipation scale
given by $\W$, the cut-off frequency. The non-dissipative limit in this
situation corresponds to taking this dissipative cut-off to infinity. 
Figs. \ref{fig-infiniteCapacity} and \ref{fig-finiteCapacity} show the
total dissipated energy as a function of time for two decay systems with the 
same initial condition,  $\N{\w}(0)=\delta(\w-1)$, interacting via the
product kernel, Eq.~(\ref{eq-productKernel}) with two different values
of $\lambda$. Fig.~\ref{fig-infiniteCapacity} shows total dissipated energy
as a function of time for the case $\lambda=3/4$ which is infinite capacity.
As the cut-off, $\W$, is increased the time at which the energy is dissipated
increases. Consequentially, the dissipated energy at any fixed time vanishes
as the dissipative cut-off is removed to infinity. Hence there is no
dissipative anomaly. The corresponding situation for the finite capacity
case, $\lambda=3/2$, shown in Fig.~\ref{fig-finiteCapacity} is strikingly
different. As the dissipative cut-off, $\W$, is removed, the dissipated energy
as a function of time becomes {\em independent} of $\W$. From the figure it
is clear that for times larger than about 3, the dissipated energy is finite
in the limit $\w\to\infty$. This is as explicit  a demonstration as one may
expect from a numerical simulation of the presence of a dissipative anomaly in 
this system. 

It is worth noting that the corresponding dissipative anomaly for the 
Smoluchowski kinetic equation is well known in the aggregation literature 
where it is more commonly referred to as the {\em gelation transition}. The
criterion, $\lambda>1$, is common to both systems. A numerical study 
similar to what has been presented here is provided for that system in 
\cite{CRZ2009}. An exact solution for the Smoluchowski equation with the
product kernel which exhibits the anomaly explicitly can be found in 
Sec. 4.5 of \cite{LEY2003}. In the light of the numerical results presented
here, the construction of the corresponding mathematically rigorous solution
of the 3-wave kinetic equation would likely be a fruitful line of research.

\section{The Bottleneck Effect in Systems With Open Truncation}
\label{sec-bottleneck}
\begin{figure}
\includegraphics[width=6.5cm]{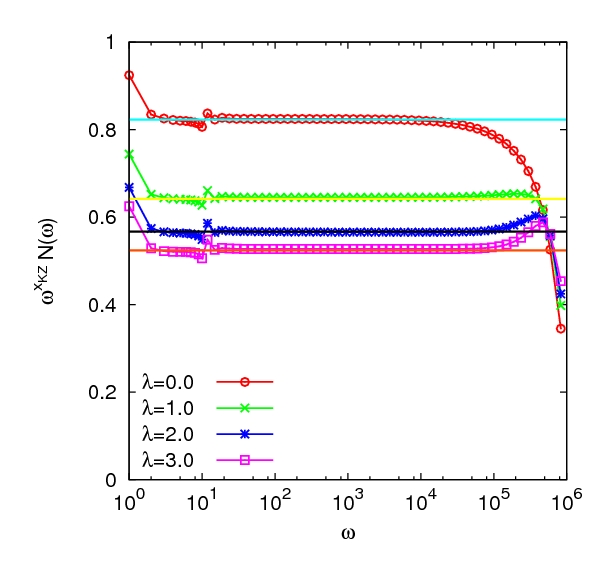}\\
\caption{\label{fig-bottleneck} The bottleneck effect in the 3-wave
kinetic equation with constant energy injection at $\w=1$ and open truncation
at $\W=10^6$. The plot shows stationary spectra compensated by the K-Z
scaling for the product kernel, Eq.~(\ref{eq-productKernel}), for a 
range of values of $\lambda$. The bottleneck effect results in
an accumulation or depletion of the spectrum near the cut-off, depending on
the choice of kernel. Solid lines indicate the theoretically predicted
values of the K-Z constant from Eq.~(\ref{eq-KZconstant}).}
\end{figure}

In the previous section, we were interested in the bahaviour of the solution 
of the 3-wave
kinetic equation with open truncation at fixed time as the dissipative 
cut-off, $\W$, tends to infinity.  In this section, we shall consider the 
complementary situation where $\W$ is fixed and time tends to infinity. In this
situation, irrespective of whether there is a dissipative anomaly or not, we
always expect the solution to tend to a stationary state corresponding to the 
Kolmogorov--Zakharov spectrum, Eq.~(\ref{eq-KZSpectrumw}), modified by the 
presence of the finite cut-off. A question of interest is how the K-Z spectrum
matches to the dissipation range and the first issue which arises is whether
or not there is a ``bottleneck'' effect.

The term ``bottleneck'' refers to a phenomenon, initially observed in numerical
simulations of Navier--Stokes turbulence (see \cite{DHYA2003} and the
references therein) where
the stationary spectrum exhibits a ``bump'' super-imposed upon the expected
constant flux spectrum as it enters the dissipation range. 
A pysical mechanism for the bottleneck
was suggested in \cite{FAL1994}. A corresponding theory for the wave
turbulence case was suggested in \cite{FR1990}. 
It is an intrinsically dissipative phenomenon. It arises because the rate of 
forward transfer of energy from frequency $\w_1$ due to the interaction with
frequency $\w_2 > \w_1$ is proportional to the product, $\N{\w_1}\N{\w_2}$, of 
occupation numbers of both frequencies (see Eq.~(\ref{eq-S1Rates})). 
The $\w_2$ contributing to the total rate of transfer of energy can be divided
into those $\w_2$ less than the dissipative scale, $\W$, and those greater
than $\W$. The latter set are in the dissipative range, where $\N{\w_2}$ is 
effectively zero (exactly zero in the case of dissipation via an open 
truncation as is considered in this article). Therefore the rate of forward 
transfer of energy through $\w_1$ would be decreased. However since $\w_1$
is still in the inertial range, the rate of energy transfer must remain equal
to the injected flux. Hence the occupation numbers of those $\w_2$ between
$\w_1$ and $\W$ must increase so that the same flux can be carried by fewer
triads (hence the term ``bottleneck''). This effect then produces the bump at 
the end of the spectrum.

Here we investigate the bottleneck effect in the 3-wave kinetic equation 
explicitly using the numerical algorithm developed in Sec.~\ref{sec-numerics}.
Fig.~\ref{fig-bottleneck} shows some results. We computed the stationary 
state of Eq.~(\ref{eq-3WKEB}) with an open truncation at $\W=10^6$ and the
product kernel, Eq.~(\ref{eq-productKernel}) for several different values
of $\lambda$. Fig.~\ref{fig-bottleneck} shows several such stationary spectra
compensated by the corresponding K-Z spectra. The solid lines indicate the
fitted values of the K-Z constant used for validation of the code in
Sec.~\ref{sec-validation}. We see that there is non-trivial structure 
super-imposed upon the K-Z spectrum as it approaches the dissipative cut-off.
Interestingly, whether this structure corresponds to a bump or not depends
upon the value of $\lambda$. The heuristic argument outlined above might
lead one to expect that the bottleneck effect always produces a bump whereas,
in reality, this does not seem to be the case. From the numerical 
calculations, it seems that the matching of the spectrum to the dissipation
produces a bump for $\lambda >1$ but produces a hollow for $\lambda<1$. We
shall remain agnostic on the question of whether it is appropriate to call
each of these regimes a bottleneck. While this is not a systematic study,
it suggests that it might be worthwhile to revisit the bottleneck phenomenon
in wave turbulence.

\section{Thermalisation in Systems with Closed Truncation}
\label{sec-thermalisation}
\begin{figure*}
\begin{center}
\begin{tabular}{cc}
\includegraphics[width=6.5cm]{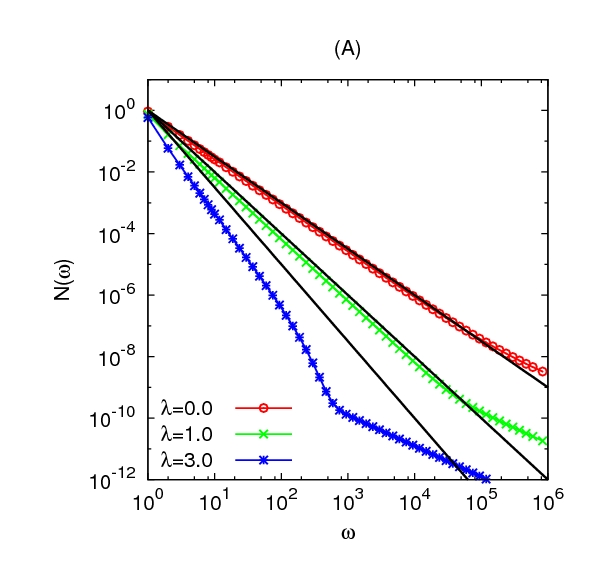}&
\includegraphics[width=6.5cm]{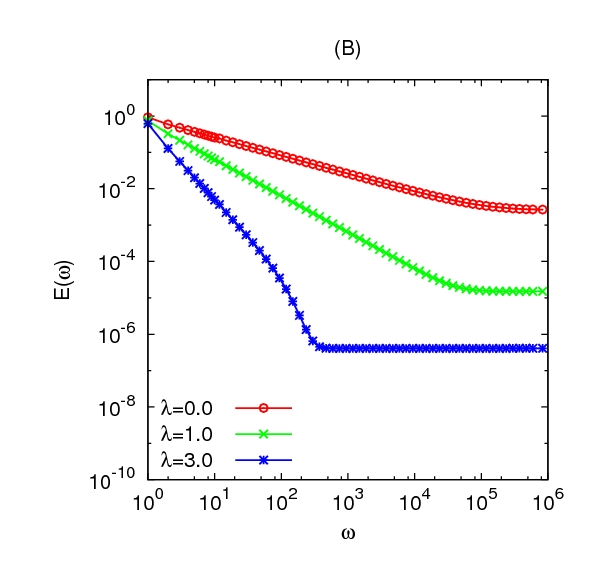}
\end{tabular}
\end{center}
\caption{\label{fig-thermalisation} Thermalisation in decay simulations of
the 3-wave kinetic equation, Eq.~(\ref{eq-3WKEB}),  with closed truncation
($\gamma=0$)  at $\W=10^6$ for the
product kernel, Eq.~(\ref{eq-productKernel}), for several different
values of $\lambda$. The left panel shows the wave spectrum, $\N{\w}$, and
the right panel shows the corresponding energy distribution, $E_\w$,
from which the equipartition effect near the cut-off is clear.}
\end{figure*}

\begin{figure}
\includegraphics[width=6.5cm]{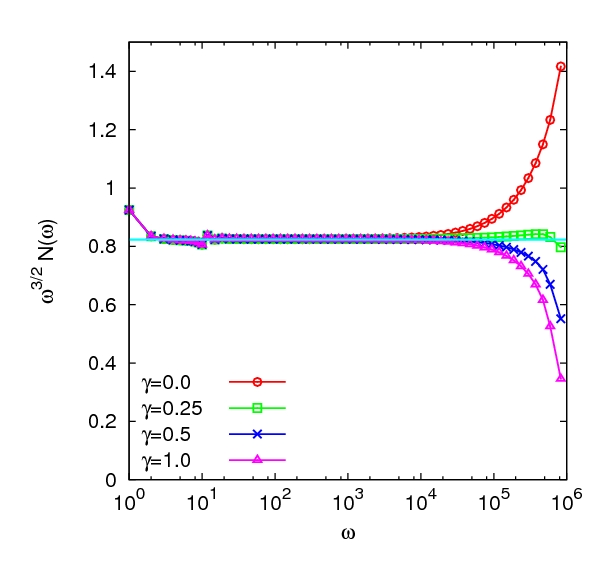}\\
\caption{\label{fig-partial_thermalisation} Partial thermalisation in the
3-wave kinetic equation, Eq.~(\ref{eq-3WKEB}), with constant kernel and a partially open 
truncation at $\W=10^6$. The plot shows stationary spectra (with the
exception of the case $\gamma=0$ for which there is no stationary state)
compensated by the K-Z scaling $\w^{3/2}$ for several different values of
$\gamma$. }
\end{figure}

In the previous two sections, we investigated two phenomena - the dissipative
anomaly and the bottleneck effect - which were related to the choice of
open truncation. In this section we consider the 3-wave kinetic equation
with closed truncation where a new phenomenon arises known as thermalisation.

We have already seen that Eq.~(\ref{eq-3WKE}) has two stationary scaling 
solutions - a thermodynamic solution and a KZ solution. The former has
finite temperature and zero energy flux whereas the latter has finite
energy flux and zero temperature. It is conjectured \cite{ZLF92,DNPZ92} that
the general solution of  Eq.~(\ref{eq-3WKE}) should be a two-parameter
function depending on both a flux and a temperature although little is
known about what such mixed states should look like except, perhaps,
as pertubations of the pure K-Z or pure thermodynamic solutions 
(see chap. 4 of \cite{ZLF92}). They were first realised
in \cite{CN04} in the context of the Leith model - a very simplified model
of hydrodynamic turbulence - and soon after were properly observed in the full 
Euler equations \cite{BBDB2005}. Since then, considerable effort has been 
invested in understanding the interplay between thermalisation and
turbulence in the hydrodynamic context 
\cite{FK2008,BB2006,KMBP2008} where it is argued \cite{FK2008} that the
closed truncation arises naturally. Despite all of this activity, the
phenomenon has not yet been investigated in the context of wave turbulence. 

In order for thermalisation to be possible, the energy of the truncated system 
should be conserved. Following the discussion of Sec.~\ref{sec-truncation}, we 
should choose $\gamma=0$ in Eq.~(\ref{eq-truncatedKE}) and work with the 
closed truncation. We performed a series of decay calculations of 
Eq.~(\ref{eq-truncatedKE}) with initial condition $\N{\w}=\delta(\w-1)$
and truncation frequency $\W=10^6$ for the product kernel, 
Eq.~(\ref{eq-productKernel}), with a range of values of $\lambda$. Some 
representative results are shown in Fig.~\ref{fig-thermalisation}. Since
these are decay simulations, the spectra shown in Fig.~\ref{fig-thermalisation}
are not stationary. The spectra are presented at times for which the
corresponding system with an open truncation would have dissipated half of
the initial energy. Thus the results of Fig.~\ref{fig-thermalisation}
are directly comparable with the decay simulations used to investigate the
dissipative anomaly in Sec.~\ref{sec-dissipativeAnomaly}. 
Fig.~\ref{fig-thermalisation}(A) shows the spectra for several different 
values of $\lambda$. The solid black lines indicate the corresponding K-Z
spectra. The effect of the closed truncation is clearly evident in the increase
of the spectrum above its K-Z value near the cut-off. That this corresponds
to thermalisation of the high frequencies is evident from 
Fig.~\ref{fig-thermalisation}(B) which shows the corresponding energy
spectra. It is clear that that the spectrum is crossing over to an 
equipartition of energy near the cut-off. 
 
Unlike the dissipative bottleneck associated with the open truncation which
we studied in Sec.~\ref{sec-bottleneck}, thermalisation always leads to an 
accumulation of energy near the cut-off, irrespective of the interaction
coefficient. The distinction is clearly illustrated by comparing the 
$\lambda=0$ cases for the closed and open truncations. The closed truncation
leads to accumulation of energy near the cut-off as shown in 
Fig.~\ref{fig-thermalisation}(A) whereas the open truncation leads to
depletion of energy near the cut-off as shown in Fig.~\ref{fig-bottleneck}.
Since thermalisation always leads to accumulation of energy near the
cut-off, this effect is sometimes also referred to as a ``bottleneck''.
In the current study, such dual use of terminology would be quite confusing 
since the closed truncation leading to thermalisation has, by construction, 
zero flux at $\w=\W$. On the other hand, the open truncation leading
to the bottleneck effect has, by construction, a finite flux at $\w=\W$.

The two phenomena can be linked to each other using the partially
open truncation discussed in Sec.~\ref{sec-truncation}. Although it seems
unlikely that such a boundary condition is of much relevance to any physical
system, choosing the parameter $\gamma$ in Eq.~(\ref{eq-3WKEB}) to have
intermediate values between 0 and $1$ allows us to interpolate smoothly
between the open and closed truncation and hence between small
scale thermalisation and small scale bottleneck. The results of one such
exercise in numerical  trickery is shown in 
Fig.~\ref{fig-partial_thermalisation}. In this figure, spectra compensated
by the K-Z scaling are shown which were obtained by solving Eq.~(\ref{eq-3WKEB})
with constant kernel ($\lambda=0$) and several different values of $\gamma$ for a system truncated at 
$\W=10^6$. The spectra shown for the finite $\gamma$ are truly stationary.
Unlike the spectra in Fig.~\ref{fig-thermalisation} discussed above, the 
spectra in Fig.~\ref{fig-partial_thermalisation} were obtained with a forcing
term injecting energy at a constant rate. Obviously this was not possible
for the case $\gamma=0$ where the total energy diverges in the forced case. The spectrum
for $\gamma$ is not stationary and is displayed at a time which allows
it to be compared with the other spectra. The message to be taken from 
Fig.~\ref{fig-thermalisation} is that the depletion of energy near the cut-off 
due to the open cut-off goes over to a thermalised accumulation of energy near 
the cut-off as the efficiency of the dissipation is decreased. It would be
interesting to investigate the relationship between the open and closed
truncations more carefully, especially to understand the role played by
the interaction coefficient in determining the shape of the spectrum near
the cut-off.

\section{Conclusions}
\label{sec-conclusion}
To conclude, we have outlined an analogy between the isotropic 3-wave
kinetic equation and the rate equations for a aggregation--fragmentation
problem with an unusual nonlinear fragmentation mechanism. This analogy
demonstrates that almost all properties of the system are determined by
a single scaling parameter, $\lambda=\frac{2\beta-\alpha}{\alpha}$ thus
greatly reducing the parameter space of possible behaviours. A new numerical
scheme was constructed based on this analogy which allows for the stable
integration of the isotropic 3-wave kinetic equation over many decades of
frequency space. This algorithm was validated by comparing numerical 
measurements of the stationary state with theoretical calculations of the
K-Z constant for a range of model interaction coefficients. Several 
applications of the new algorithm were then presented including studies
of the dissipative anomaly, bottleneck effect and thermalisation phenomenon.

The preliminary results presented here by way of motivation for the study
of the isotropic 3-wave kinetic equation have suggested that some further
investigation of cut-off related phenomena such as the bottleneck would likely
be fruitful. In addition, since the initial studies of Galtier et al. 
\cite{GNNP2000} of the solutions of the Alfven wave kinetic equation, there
is growing evidence \cite{LEE2001,CNP03,CN04} that finite capacity 
cascade often, if not always, exhibit a dynamical scaling anomaly during
the transient stage of evolution which cannot be understood from elementary
scaling arguments. The methods developed in this article should provide an
ideal set of tools to study this phenomenon in the setting of general kinetic
equations. This will form the basis of future work. The other obvious line
or research which has opened up is the question of whether the methods decribed
here can be extended to the case of the isotropic 4-wave kinetic equation.
At this point, the answer does not seem obvious. The key 
approximation used in Sec.~\ref{sec-numerics} to develop the numerical
algorithm involved treating all waves in the leftmost bin as having the 
same frequency. This may not be a reasonable approximation in a 
4-wave system where the waves interact in quartets and the possibility of
an inverse cascade may increase the sensitivity of the dynamics to the
way in which low frequencies are approximated.

\section{Acknowlegements}
The author acknowledges helpful discussions with E. Ben-Naim, P. Krapivsky, 
Y. Lvov, A.C.  Newell and Y. Pomeau and thanks F. Leyvraz for bringing the work of M.H. Lee 
to his attention. 

\bibliography{all}

\section*{Appendix: Derivation of the Aggregation--Fragmentation Equations}
\label{sec-app1}
We consider isotropic wave turbulence. The wave spectrum , $\nkv{k}$, is
therefore a function of $k=\left|\k\right|$ only which we shall denote
by $\n{k}$. There exist a strightforward set of changes of variables which
take advantage of this isotropy to convert the collision integral over the 
pair of $d$--dimensional wave-vectors, $\k_1$ and $\k_2$, into one-dimensional
integrals over frequencies.

The total number of waves in the system is
\begin{displaymath}
N = \int_{{\mathbb R}^d} \nkv{k}\,d\k.
\end{displaymath}
Denoting the radial coordinate of $\k \in {\mathbb R}^d$ and integration over 
the corresponding angular variables by $\dTd{k}$  we can write this integral
in spherical polar co-ordinates as:
\begin{eqnarray*}
N&=&\int_0^\infty k^{d-1}\, dk \int \dTd{k}   \nkv{k} \\ 
&=&  \Wd \int_0^\infty \n{k} k^{d-1} dk
\end{eqnarray*}
where we have used the isotropy of $\nkv{k}$ to integrate over the
angular variables and denoted the resulting $d$-dimensional solid angle by 
$\Wd$. We now use the isotropy of the dispersion relation, 
\begin{displaymath}
\w(\k) = c\,k^\alpha
\end{displaymath}
to change variables from integration over $k$ to integration over
$\w$:
\begin{eqnarray}
\nonumber N &=&  \Wd \int_0^\infty \n{k} k^{d-1} \dd{k}{\w} d\w\\
\label{eq-transformToFreq}&=& \frac{\Wd}{\alpha}\, c^{-\frac{\alpha}{d}} \int_0^\infty \n{\w}\,\w^{\frac{d-\alpha}{\alpha}}\,d\w.
\end{eqnarray}
where $\n{\w} = \n{k(\w)}$.
Based on these manipulations,  we define the angle-averaged frequency 
spectrum, $\N{\w}$, by
\begin{equation}
\N{\w} = \frac{\Wd}{\alpha}\, c^{-\frac{\alpha}{d}} \w^{\frac{d-\alpha}{\alpha}}\,\n{\w}
\label{eq-Nw}
\end{equation}
The angle-averaged frequency spectrum has the advantage that the total
number of waves, $N$, and total wave energy, $E$, are given very simply as
\begin{eqnarray}
N&=&\int_0^\infty \N{\w}\,d\w\\
E&=&\int_0^\infty \w \N{\w}\,d\w.
\end{eqnarray}
Our objective is to integrate over angular variables and express the
kinetic equation entirely in terms of $\N{\w}$.

We begin by using Eq.~(\ref{eq-3WKE}), to write an evolution equation for the time evolution of $\N{\w_1}$:
\begin{eqnarray*}
\label{eq-SSplit1}\int_0^\infty \pd{\N{\w_1}}{t} d\w_1 &=& \int_{\mathbb R}^d S[\nkv{k}]\,d\k_1\\
\nonumber&\equiv& \int_0^\infty (S_1[\N{\w}] + S_2[\N{\w}] + S_3[\N{\w}] )\, d\w_1.
\end{eqnarray*}
This yields a kinetic equation of the form
\begin{equation}
\pd{\N{\w_1}}{t} = S_1[\N{\w}] + S_2[\N{\w}] + S_3[\N{\w}]. 
\end{equation}
To determine the form of $S_1[\N{\w}]$, $S_2[\N{\w}]$ and $S_3[\N{\w}]$
we split the collision integral, $S[\nkv{k}]$, into three terms, as follows:
\begin{eqnarray}
\label{eq-SSplit2} & &\int_{{\mathbb R}^d} S[n_{\k}] d\k_1 = a_1 T_1 + a_2 T_2 + a_3 T_3\\
 \nonumber &=& 4\pi \int L_{\k_1\k_2\k_3}^2 (a_1 n_{\k_2} n_{\k_3} -  a_2 n_{\k_1}n_{\k_3}- a_3 n_{\k_1}n_{\k_2}) \\
\nonumber && \hspace{0.5cm} \delta(\omega_{\k_1}-\omega_{\k_2}-\omega_{\k_3})\, \delta(\k_1-\k_2-\k_3)\, d\k_1d\k_2d\k_3 \\
\nonumber &-&4\pi \int L_{\k_2\k_3\k_1}^2 (a_1 n_{\k_1}n_{\k_3} -  a_2 n_{\k_2}n_{\k_3}- a_3 n_{\k_1}n_{\k_2}) \\
&&\nonumber \hspace{0.5cm}  \delta(\omega_{\k_2}-\omega_{\k_3}-\omega_{\k_1})\, \delta(\k_2-\k_3-\k_1)\, d\k_1d\k_2d\k_3 \\
\nonumber &-&4 \pi \int L_{\k_3\k_1\k_2}^2 (a_1 n_{\k_1}n_{\k_2} -  a_2 n_{\k_1}n_{\k_3}+ a_3 n_{\k_2}n_{\k_3}) \\
\nonumber && \hspace{0.5cm} \delta(\omega_{\k_3}-\omega_{\k_1}-\omega_{\k_2})\, \delta(\k_3-\k_1-\k_2)\, d\k_1d\k_2d\k_3
\end{eqnarray}
The variables $a_1 = a_2 = a_3 = 1$ have been introduced simply to indicate
which terms are to be grouped together.

Let us first consider the terms proportional to $a_1$. We again introduce
spherical polar coordinates: 
\begin{equation}
\int d\k_1 d\k_2 d\k_3 = \int (k_1k_2k_3)^{d-1} \dTd{k_1}\dTd{k_2}\dTd{k_3} dk_{123},
\end{equation}
where the notation $dk_{123}$ represents the integration measure, 
$dk_1dk_2dk_3$,
over the radial variables.  Noting that the angular variables only enter into 
the interaction coefficients, $L_{\k_1\k_2\k_3}$ and the $\k$ delta functions, 
we can define an angle-averaged interaction coefficient,
\begin{eqnarray}
\label{eq-LtildeDefn}
\widetilde{L}(k_1,k_2,k_3) &=& 4\pi \int L_{\k_1\k_2\k_3}^2 \delta(\k_1\!-\!\k_2\!-\!\k_3)\\
\nonumber  & &\hspace{1.0cm} \dTd{k_1}\dTd{k_2}\dTd{k_3},
\end{eqnarray}
which is a function of the radial variables only. We can then write
\begin{eqnarray}
\nonumber T_1 &=& \int \widetilde{L}_{k_1k_2k_3}\, \n{k_2} \n{k_3} (k_1k_2k_3)^{d-1}\, \delta(\w^1_{23})\, dk_{123}\\
\nonumber &-& \int \widetilde{L}_{k_2k_3k_1}\, \n{k_1} \n{k_3} (k_1k_2k_3)^{d-1}\, \delta(\w^2_{31})\, dk_{123}\\
\nonumber &-& \int \widetilde{L}_{k_3k_1k_2}\, \n{k_1} \n{k_2} (k_1k_2k_3)^{d-1}\, \delta(\w^3_{12})\, dk_{123}.
\end{eqnarray}
Here the notation $\delta(\w^i_{jk})$ is a compact representation of the
frequency delta function:
\begin{equation}
\delta(\w^i_{jk}) = \delta(\w_i-\w_j-\w_k).
\end{equation}
Now replace the integration over $k$'s with integration over frequencies as
we did in Eq.~(\ref{eq-transformToFreq}) and
use Eq.~(\ref{eq-Nw}) to express $\n{\w}$ in terms of $\N{\w}$. The result
is
\begin{eqnarray}
\nonumber T_1 &=& c_d \int \bar{L}_{\w_1\w_2\w_3}\,\w_1^{\frac{d-\alpha}{\alpha}}\, \N{\w_2} \N{\w_3} \delta(\w^1_{23})\, d\w_{123}\\
\label{eq-T1B} &-& c_d \int \bar{L}_{\w_2\w_3\w_1}\,\w_2^{\frac{d-\alpha}{\alpha}}\, \N{\w_1} \N{\w_3} \delta(\w^2_{31})\, d\w_{123}\\
\nonumber  &-& c_d \int \bar{L}_{\w_3\w_1\w_2}\,\w_3^{\frac{d-\alpha}{\alpha}}\, \N{\w_1} \N{\w_2} \delta(\w^3_{12})\, d\w_{123},
\end{eqnarray}
where
\begin{eqnarray}
\label{eq-LbarDefn}
\bar{L}_{\w_1\w_2\w_3} &=&  \widetilde{L}_{\left(\frac{\w_1}{c}\right)^\frac{1}{\alpha} \left(\frac{\w_2}{c}\right)^\frac{1}{\alpha} \left(\frac{\w_3}{c}\right)^\frac{1}{\alpha}}\\
\nonumber c_d &=&\frac{c^{\frac{4-d}{\alpha}}}{\alpha {\Wd}^2}.
\end{eqnarray}
Finally define
\begin{equation}
\label{eq-KDefn}
K_1(\w_i,\w_j) = c_d  \bar{L}_{\w_i+\w_j\,\w_i\w_j}\,(\w_i+\w_j)^{\frac{d-\alpha}{\alpha}}.
\end{equation}
Now use the frequency delta-functions to write Eq~(\ref{eq-T1B}) as
\begin{eqnarray}
\nonumber T_1 &=& \int K_1(\w_2,\w_3)\, \N{\w_2} \N{\w_3} \delta(\w^1_{23})\, d\w_{123}\\
\label{eq-T1C} &-& \int K_1(\w_3,\w_1)\, \N{\w_1} \N{\w_3} \delta(\w^2_{31})\, d\w_{123}\\
\nonumber  &-& \int K_1(\w_1,\w_2)\,\N{\w_1} \N{\w_2} \delta(\w^3_{12})\, d\w_{123},
\end{eqnarray}
Comparing with Eq.~(\ref{eq-SSplit1}) we see that we should write
\begin{eqnarray}
\nonumber S_1[\N{\w}]\!\!\! &=& \!\!\!\!\!\int\!\! K_1(\w_2,\w_3)\, \N{\w_2} \N{\w_3} \delta(\w_1\!-\!\w_2\!-\!\w_3)\, d\w_{23}\\
\label{eq-S1B} &-& \!\!\!\!\!\int\!\! K_1(\w_3,\w_1)\, \N{\w_1} \N{\w_3} \delta(\w_2\!-\!\w_3\!-\!\w_1)\, d\w_{23}\\
\nonumber  &-& \!\!\!\!\!\int\!\! K_1(\w_1,\w_2)\,\N{\w_1} \N{\w_2} \delta(\w_3\!-\!\w_1\!-\!\w_2)\, d\w_{23}.
\end{eqnarray}
The same set of manipulations can now be applied to the terms proportional
to $a_2$ and $a_3$ in Eq.~(\ref{eq-SSplit2}) to deduce the appropriate forms
of $S_2[\N{\w}]$ and $S_3[\N{\w}]$. The results are as follows:
\begin{eqnarray}
\nonumber S_2[\N{\w}]\!\!\! &=& \!\!\!\!\!\int\!\! K_2(\w_2,\w_3)\, \N{\w_1} \N{\w_3} \delta(\w_1\!-\!\w_2\!-\!\w_3)\, d\w_{23}\\
\label{eq-S2B} &-& \!\!\!\!\!\int\!\! K_2(\w_3,\w_1)\, \N{\w_2} \N{\w_3} \delta(\w_2\!-\!\w_3\!-\!\w_1)\, d\w_{23}\\
\nonumber  &-& \!\!\!\!\!\int\!\! K_2(\w_1,\w_2)\,\N{\w_1} \N{\w_3} \delta(\w_3\!-\!\w_1\!-\!\w_2)\, d\w_{23}
\end{eqnarray}
and
\begin{eqnarray}
\nonumber S_3[\N{\w}]\!\!\! &=& \!\!\!\!\!\int\!\! K_3(\w_2,\w_3)\, \N{\w_1} \N{\w_2} \delta(\w_1\!-\!\w_2\!-\!\w_3)\, d\w_{23}\\
\label{eq-S3B} &-& \!\!\!\!\!\int\!\! K_3(\w_3,\w_1)\, \N{\w_1} \N{\w_2} \delta(\w_2\!-\!\w_3\!-\!\w_1)\, d\w_{23}\\
\nonumber  &-& \!\!\!\!\!\int\!\! K_3(\w_1,\w_2)\,\N{\w_2} \N{\w_3} \delta(\w_3\!-\!\w_1\!-\!\w_2)\, d\w_{23}.
\end{eqnarray}
Note that there is a small price to be paid for hiding all dependence on $d$
and $\alpha$: the interaction coefficients for the three collision
integrals are, in general, not identical:
\begin{eqnarray}
\nonumber K_2(\w_i,\w_j) &=& K_1(\w_i,\w_j) \left(\frac{\w_i+\w_j}{\w_j}\right)^{\frac{\alpha-d}{\alpha}}\\
\label{eq-defnK2K3}K_3(\w_i,\w_j) &=& K_1(\w_i,\w_j) \left(\frac{\w_i+\w_j}{\w_i}\right)^{\frac{\alpha-d}{\alpha}}.
\end{eqnarray}
Note also that $K_2(\w_i,\w_j)$ and $K_3(\w_i,\w_j)$, are not symmetric in 
their arguments although this latter deficiency can be removed if desired
by symmetrisation. These problems are immaterial since the original variables
can be easily restored if needs be. It is worth noting that in the case where
$d=\alpha$, the distinctions between the interaction coefficients disappear.
Furthermore, even in the general case, all three interaction coefficients
have the same degree of homogeneity. From Eq.~(\ref{eq-KDefn}), 
Eq.~(\ref{eq-LbarDefn}) and Eq.~(\ref{eq-LtildeDefn}), it is easy to work
backwards and establish that this degree of homogeneity, which we
denote by $\lambda$,  is
\begin{equation}
\lambda=\frac{2\beta - \alpha}{\alpha}.
\end{equation}
Thus, if one is interested in scaling properties of 3-wave kinetic equation,
almost everything is determined by this single scaling parameter, $\lambda$.

\section*{Appendix: Transfer integrals}
\label{sec-app2}
In this appendix we state the explicit expressions for the various
transfer integrals which go into the estimation of the collision integral
according to Eq.~(\ref{eq-CaseB}) and Eq.~(\ref{eq-CaseC}).

\begin{itemize}
\item{{\bf Case B} :\\}
\begin{eqnarray*}
^{B}L_j^{(1)} &=& \W_j\,N_j \int_{\w_k^L}^{\w_k^R} K_1(\W_j,\w_k)\,n_k(\w_k)\,dw_k\\
^{B}L_k^{(1)} &=& N_j \int_{\w_k^L}^{\w_k^R} \w_k\,K_1(\W_j,\w_k)\,n_k(\w_k)\,dw_k\\
^{B}G_j^{(2)} &=& \W_j\,N_j \int_{\w_k^L}^{\w_k^R} K_2(\W_j,\w_k)\,n_i(\W_j+\w_k)\\
^{B}G_k^{(2)} &=& N_j \int_{\w_k^L}^{\w_k^R} \w_k\,K_2(\W_j,\w_k)\,n_i(\W_j+\w_k)\\
^{B}G_j^{(3)} &=& \W_j\,\Delta\w_j \int_{\w_k^L}^{\w_k^R} K_3(\W_j,\w_k)\,n_k(\w_k)\,n_i(\W_j+\w_k)\\
^{B}G_k^{(3)} &=& \Delta\w_j \int_{\w_k^L}^{\w_k^R} \w_k\,K_3(\W_j,\w_k)\,n_k(\w_k)\,n_i(\W_j+\w_k)\\
\end{eqnarray*}
\item{{\bf Case C} :\\}
\begin{eqnarray*}
_L^{C}L_j^{(1)} &=& \W_j\,N_j \int_{\w_k^L}^{\W^*} K_1(\W_j,\w_k)\,n_k(\w_k)\,dw_k\\
_L^{C}L_k^{(1)} &=& N_j \int_{\w_k^L}^{\W^*} \w_k\,K_1(\W_j,\w_k)\,n_k(\w_k)\,dw_k\\
_L^{C}G_j^{(2)} &=& \W_j\,N_j \int_{\w_k^L}^{\W^*} K_2(\W_j,\w_k)\,n_{i_L}(\W_j+\w_k)\\
_L^{C}G_k^{(2)} &=& N_j \int_{\w_k^L}^{\W^*} \w_k\,K_2(\W_j,\w_k)\,n_{i_L}(\W_j+\w_k)\\
_L^{C}G_j^{(3)} &=& \W_j\,\Delta\w_j \int_{\w_k^L}^{\W^*} K_3(\W_j,\w_k)\,n_k(\w_k)\,n_{i_L}(\W_j+\w_k)\\
_L^{C}G_k^{(3)} &=& \Delta\w_j \int_{\w_k^L}^{\W^*} \w_k\,K_3(\W_j,\w_k)\,n_k(\w_k)\,n_{i_L}(\W_j+\w_k)\\
\end{eqnarray*}
\begin{eqnarray*}
_R^{C}L_j^{(1)} &=& \W_j\,N_j \int_{\W^*}^{\w_k^R} K_1(\W_j,\w_k)\,n_k(\w_k)\,dw_k\\
_R^{C}L_k^{(1)} &=& N_j \int_{\W^*}^{\w_k^R} \w_k\,K_1(\W_j,\w_k)\,n_k(\w_k)\,dw_k\\
_R^{C}G_j^{(2)} &=& \W_j\,N_j \int_{\W^*}^{\w_k^R} K_2(\W_j,\w_k)\,n_{i_R}(\W_j+\w_k)\\
_R^{C}G_k^{(2)} &=& N_j \int_{\W^*}^{\w_k^R} \w_k\,K_2(\W_j,\w_k)\,n_{i_R}(\W_j+\w_k)\\
_R^{C}G_j^{(3)} &=& \W_j\,\Delta\w_j \int_{\W^*}^{\w_k^R} K_3(\W_j,\w_k)\,n_k(\w_k)\,n_{i_R}(\W_j+\w_k)\\
_R^{C}G_k^{(3)} &=& \Delta\w_j \int_{\W^*}^{\w_k^R} \w_k\,K_3(\W_j,\w_k)\,n_k(\w_k)\,n_{i_R}(\W_j+\w_k)\\
\end{eqnarray*}

\end{itemize}

\end{document}